\documentclass[english,reprint,nofootinbib]{revtex4-1}
\pdfoutput=1
\usepackage[T1]{fontenc}
\usepackage[latin9]{inputenc}
\usepackage{geometry}
\usepackage{xcolor}
\usepackage{babel}
\usepackage{verbatim}
\usepackage{float}
\usepackage{amsmath}
\usepackage{amssymb}
\usepackage{graphicx}
\usepackage{esint}
\usepackage[unicode=true]
 {hyperref}

\makeatletter

\providecommand{\tabularnewline}{\\}

\makeatother

\begin{document}
\title{Dynamical persistence in resource-consumer models}
\author{Itay Dalmedigos}
\affiliation{Department of Physics, Technion-Israel Institute of Technology, Haifa
32000, Israel}
\author{Guy Bunin}
\affiliation{Department of Physics, Technion-Israel Institute of Technology, Haifa
32000, Israel}
\begin{abstract}
We show how highly-diverse ecological communities may display persistent
abundance fluctuations, when interacting through resource competition
and subjected to migration from a species pool. This turns out to
be closely related to the ratio of realized species diversity to the
number of resources. This ratio is set by competition, through the
balance between species being pushed out and invading. When this ratio
is smaller than one, dynamics will reach stable equilibria. When this
ratio is larger than one, fixed-points are either unstable or marginally
stable, as expected by the competitive exclusion principle. If they
are unstable, the system is repelled from fixed points, and abundances
forever fluctuate. While marginally-stable fixed points are in principle
allowed and predicted by some models, they become structurally unstable
at high diversity. This means that even small changes to the model,
such as non-linearities in how resources combine to generate species'
growth, will result in persistent abundance fluctuations.
\end{abstract}
\maketitle

\section{Introduction}

\global\long\def\mean{\operatorname{mean}}%
\global\long\def\var{\operatorname{var}}%
\global\long\def\std{\operatorname{std}}%
\global\long\def\corr{\operatorname{corr}}%
\global\long\def\cov{\operatorname{cov}}%
\global\long\def\tanh{\operatorname{tanh}}%

\begin{comment}
{[}{[}Itay, ECOLOGY HOMEWORK:

{*}in microbial communities, find refs: how many molecules are exchanged
between microbes? - In the 'web of microbs' examples with \textasciitilde 100-140
exchanged nutrient molecules.{]}{]}
\end{comment}

Resource competition is one of the main mechanisms underlying species
interactions. Theoretical works \citep{armstrong_competitive_1980,huisman_biodiversity_1999}
have demonstrated that communities interacting via resource competition
may exhibit different dynamical behaviors (also observed in nature
\citep{turchin_complex_1992,beninca_chaos_2008}), including relaxation
to equilibria, limit cycles and chaotic dynamics. In systems consisting
of a few species and resources, the dynamical outcome may depend on
all the details describing the interactions in the community \citep{schippers_does_2001}.
For systems with higher dimensionality (more species and resources),
full detailed knowledge of the interactions may be difficult to obtain
and predictions might seem hopeless, and potentially sensitive to
all unknown details.

In recent years, research on high-dimensional communities has shown
that full knowledge on all the interactions might not always be needed
\citep{opper_phase_1992,yoshino_statistical_2007,kessler_generalized_2015,bunin_ecological_2017,tikhonov_collective_2017,barbier_generic_2018,biroli_marginally_2018,advani_statistical_2018,roy_numerical_2019},
and important ecological quantities such as total biomass and diversity
can be predicted from a handful of statistics on the interaction parameters.
In the space of these relevant statistics, one can identify different
regions (known as ``phases'') with qualitatively distinct behaviors,
such as relaxation to equilibria versus chaotic dynamics. Within these
phases, the qualitative behavior is robust, i.e. insensitive to sufficiently
small changes in the systems' interaction coefficients.

In this paper, we consider high-dimensional communities with resource-competition
interactions. We show that in an entire region of parameter space,
the system fails to reach equilibria and instead abundances fluctuate
indefinitely. This might seem surprising, as some theoretical models
are known to always lead to stable equilibria, including classical
models by MacArthur \citep{mac_arthur_species_1969,macarthur_species_1970}.
We argue that when the number of species and resources is large, there
are regions of parameter space where these models are highly sensitive
(structurally unstable), and even very small changes to the model
will result in persistent abundance fluctuations.

A key ingredient in our discussion is competitive exclusion, according
to which the number of species that can coexist in a stable equilibrium
is smaller or equal to the number of resources (or more generally,
the number of niches). A \emph{marginally}-stable fixed point \emph{can}
accommodate more species than resources, but it can be destroyed by
small perturbations or changes to the dynamical rules.

The sensitivity of marginally-stable equilibria raises the following
question: what then replaces the marginally stable fixed-point, once
it is no longer stable? There are two possible scenarios: (1) Species
will go extinct until an equilibrium with fewer species is reached,
which satisfies the competitive exclusion principle, or (2) The system
will not reach any fixed point, and instead abundances will continue
to fluctuate indefinitely. We show that for a community experiencing
migration from a species pool, the generic situation is number (2)
above.

Resource competition dynamics in diverse communities have been analyzed
in a number of works employing tools from statistical physics \citep{yoshino_statistical_2007,tikhonov_collective_2017,advani_statistical_2018,cui_effect_2019}.
For a region of model parameters, marginally-stable \citep{tikhonov_collective_2017,landmann_systems_2018}
or close to marginally-stable \citep{yoshino_statistical_2007,advani_statistical_2018}
equilibria are reached. Yet the models studied all admit a unique
equilibrium by construction (in the spirit of classical works \citep{mac_arthur_species_1969,macarthur_species_1970}).
For example, species' growth rates are assumed to depend linearly
on resource availability, which cannot accommodate effects such as
essential resources \citep{leon_competition_1975}. A model combining
resource-competition with other interactions not mediated by resources,
was studied in \citep{yoshino_statistical_2007}. It showed that a
unique stable equilibrium cannot exist in a certain region of parameter
space, but did not study what replaces that unique equilibrium. Additional
factors might drive communities to marginal stability, such as metabolic
trade-offs \citep{posfai_metabolic_2017} or evolution, highlighting
the importance of studying the generic dynamics in these situations.

\begin{figure}
\begin{centering}
\includegraphics[width=1\columnwidth]{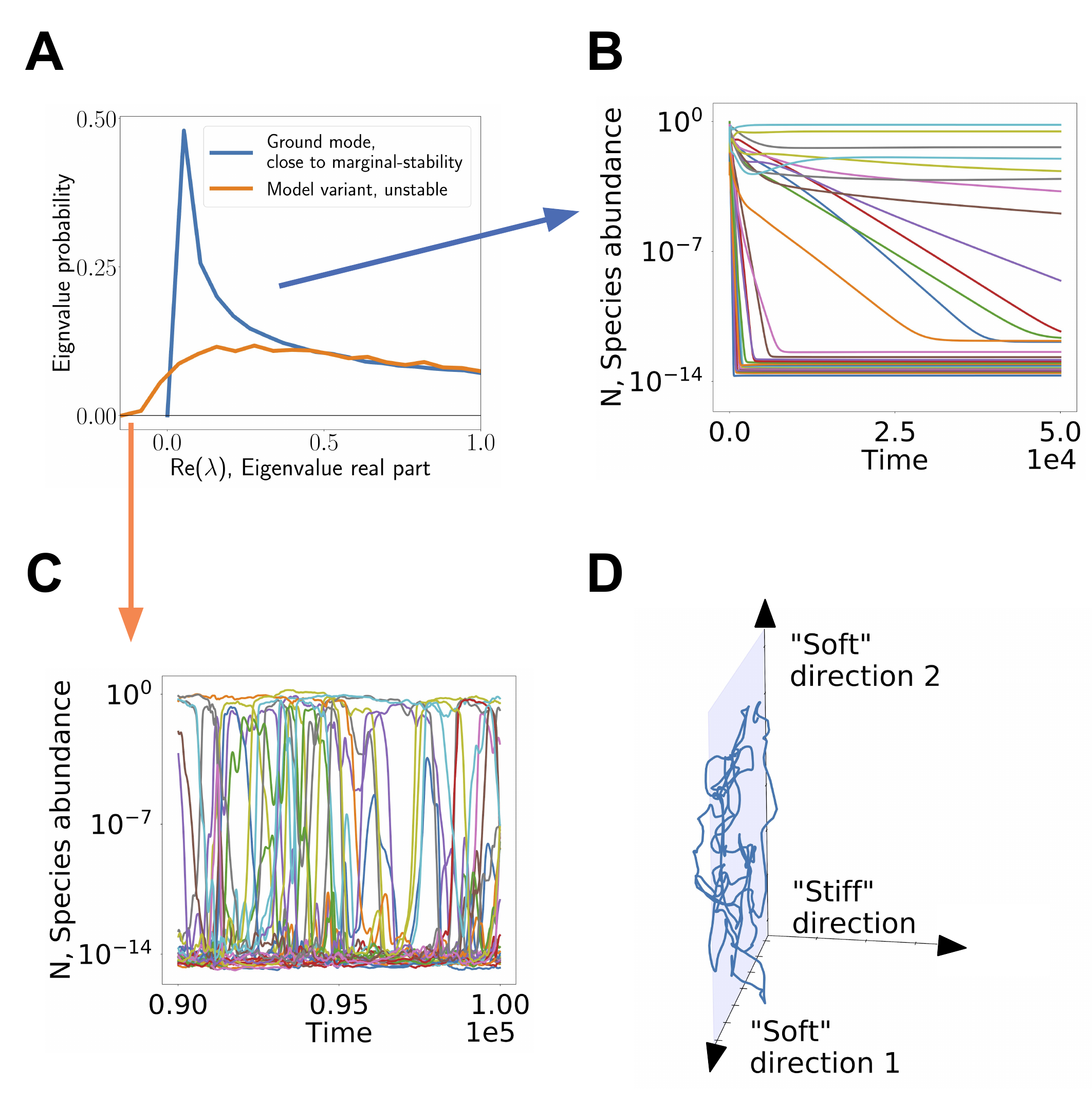}
\par\end{centering}
\caption{\label{fig:Summary-of-argument}Summary of argument. (A) The fixed
points encountered by a high-diversity resource-competition community
may be unstable. In the presence of migration this creates persistent
abundance fluctuations, shown in (C), in which species are pushed
out due to the instability, but are later able to invade again. (B)
Marginally-stable fixed points (or nearly marginal ones) that appear
in certain models, are characterized by a non-negative spectrum, also
shown in (A). They allow the community to relax to a fixed point.
But the stability of such fixed points is sensitive to modeling assumptions,
including additional interactions of other types, or how growth rates
depend on resource availability. Introducing these will generically
push the system towards unstable non-equilibrium dynamics shown in
(C). (D) In such a case, the directions corresponding to the nearly
marginal eigenvectors become ``soft'' directions, showing large
fluctuations. For clarity, in (B,C) 30 representative species are
plotted.}
\end{figure}

Our argument proceeds as follows. Interactions create a balance between
species being pushed out due to competition, and species invading
when they can, steering the community towards some target species
richness. If this richness is larger than the number of resources,
then fixed points generically will be unstable, and persist abundance
fluctuations will ensue, See Fig. \ref{fig:Summary-of-argument}(C).
These dynamics are characterized by species being pushed out by fixed
points' instability, and back when they are able to invade.

This instability is manifested by the spectrum of response to small
perturbations around a putative fixed point at the target species
richness, Fig. \ref{fig:Summary-of-argument}(A). Under certain modeling
assumptions, these fixed points might be marginally stable, but in
this case small changes to the model push the fixed point to become
truly unstable, without changing much the target richness set by the
competition, see Fig. \ref{fig:Summary-of-argument}(C). In other
words, it is precisely the large number of (nearly-)marginal directions
that allows for such fluctuating dynamics to persist, as shown Fig.
\ref{fig:Summary-of-argument}(D). Marginal, or nearly-marginal eigenvectors
around the fixed point become ``soft'' directions, namely combined
abundance fluctuations of multiple species that are met with little
resistance. This correspondence is further explored in Appendix \ref{sec:Stiff-and-soft}.

The paper is structured as follows. Sec. \ref{subsec:The-unperturbed-model}
defines the ground model used to illustrate the arguments. Sec. \ref{subsec:Sensitivity-to-direct}
looks at the effect of changes to the model, by adding interactions
on top of resource competition. It shows how the dynamics generated
by this model might vary significantly due to even small changes,
replacing equilibria by non-equilibrium dynamics. The general mechanism
behind this sensitivity is explained in Sec. \ref{subsec:Theory-for-the}.
In Sec. \ref{subsec:Resource-competition-with-non-li}, the behavior
is shown to be sensitive in a second variant of the model in which
all interactions are strictly the result of resource competition,
but with non-linear resource intake.

Sec. \ref{subsec:Intra-community-=000026-diversity} describes the
resulting abundance distributions and community diversity. The non-equilibrium
coexistence of more species than there are resources or niches, is
of great interest in its own right. It has been suggested to play
a part in the resolution of the ``paradox of the plankton'' \citep{roy_towards_2007}.
In Sec. \ref{subsec:Intra-community-=000026-diversity} we consider
this question directly in a high-dimensional setting, in light of
works on high-dimensional chaos in well-mixed communities \citep{opper_phase_1992,roy_numerical_2019}
and meta-communities \citep{roy_can_2019,pearce_stabilization_2019}.
Finally, Sec. \ref{subsec:Conclusions} concludes with a discussion,
focusing on predictions for experiments and natural communities.%

\section{Methods and Results}

\subsection{The ground model\label{subsec:The-unperturbed-model}}

To illustrate the ideas we use a well-known model and introduce two
variants to that model. The canonical model is MacArthur's resource
consumer model (MCRM) \citep{mac_arthur_species_1969}, that will
be referred below as the ``ground model''. The variants introduce
small changes to its dynamical evolution.

The MCRM describes the dynamics of $S$ species abundances $N_{i}\left(i=1\ldots S\right)$
competing over $M$ types of resources $R_{\beta}\left(\beta=1\ldots M\right)$.
The MCRM system evolves according the following set of coupled differential
equations
\begin{equation}
\begin{cases}
\frac{dN_{i}}{dt}=N_{i}\left[\sum_{\beta}c_{i\beta}R_{\beta}-m_{i}\right]+\eta_{i}\\
R_{\beta}=K_{\beta}-\sum_{j}c_{j\beta}N_{j}
\end{cases},\label{eq:MCRM EOM}
\end{equation}
where $c_{i\beta}$ describes the consumption preference of species
$i$ for resource $\beta$. $m_{i}$ is a minimum maintenance cost
that must be met by species $i$ for it to grow. $K_{\beta}$ is the
carrying capacity of resource $\beta$. The first equation includes
a migration term $\eta_{i}$ from a species pool. It will taken to
be small, allowing species to invade if they have positive growth
rates. Plugging the expression for $R_{\beta}$ into the first equation
yields%
\begin{comment}
In our first demonstrations, \textcolor{purple}{Sec. {[}..{]}}, we
will simplify the situation further, following MacArthur, and assume
that the resources equilibrate faster then the species abundances,
so that $dR_{\beta}/dt$ vanish at any given time {[}{[}Itay: we talk
about setting $dR/dt=0$ to begin with, but we use the {]}{]}, allowing
one to solve for the resource variables as a function of specie abundance's.
This is not necessary for the qualitative outcomes, but simplifies
the calculations. With this assumption, $R_{\beta}=K_{\beta}-\sum_{i}N_{i}c_{i\beta}$
{[}this assumes that $R_{\beta}$ can be negative ..{]}. This therefore
removes the explicit dependence on the resource variables, and the
dynamical equations reduce to
\end{comment}

\begin{equation}
\frac{dN_{i}}{dt}=N_{i}\left[\sum_{\beta}c_{i\beta}K_{\beta}-m_{i}-\sum_{j}\alpha_{ij}N_{j}\right]+\eta_{i}\ ,\label{eq:Reduced MCRM EOM}
\end{equation}
where $\alpha_{ij}=\alpha_{ij}^{\left(r\right)}=\sum_{\beta}c_{i\beta}c_{j\beta}$,
and the superscript $\left(r\right)$ denotes resource-mediated interactions.
This equation is now in the form of generalized Lotka-Voltera equations.

\subsection{Sensitivity to direct interactions: a demonstration\label{subsec:Sensitivity-to-direct}}

A key result by MacArthur \citep{mac_arthur_species_1969} is that
the model in Eq. (\ref{eq:Reduced MCRM EOM}) exhibits globally stable
dynamics, reaching a single fixed point independently of the system's
initial conditions. In this section we show that by a small addition
of other interactions on top of the resource competition interactions
described above, the system is no longer guaranteed to approach a
fixed point. Instead, for a broad region of control parameters, the
species' abundances fluctuate indefinitely, see Fig. \ref{fig:Summary-of-argument}(C).%
\begin{comment}
This happens precisely when the system is close to, or above, the
competitive exclusion limit. In other words, under such conditions
the unperturbed model is highly sensitive (structurally unstable),
and care should be taken when interpreting its behavior.
\end{comment}

To demonstrate this phenomena we introduce the first variant of the
MCRM, which includes additional ``direct'' species interactions,
$\alpha_{ij}^{\left(d\right)}$, so that the total interaction coefficients
read $\alpha_{ij}=\alpha_{ij}^{\left(r\right)}+\omega\cdot\alpha_{ij}^{\left(d\right)}$,
with $\omega$ controlling the strength of the perturbation. These
direct interactions may come as a result of many mechanisms that lie
beyond the unperturbed MCRM. The important point will be to find when
such additional interactions have a large effect on the dynamics,
even when they are small.

To quantify the size of the perturbation, we use the ratio of the
Frobenius norms (sum of squared interaction coefficients) of the interaction
matrices, setting $\left\Vert \omega\cdot\alpha^{\left(d\right)}\right\Vert _{F}/\left\Vert \alpha^{\left(r\right)}\right\Vert _{F}=0.05$
throughout. For any given model parameters, $\omega$ is chosen satisfy
this condition, allowing for a comparison between results with different
model parameters.

The quantities $c_{i\beta},K_{\beta},m_{i},\alpha_{ij}^{\left(d\right)}$
that define the interactions are drawn at random, representing a generic
diverse community, without any additional structure beyond that already
incorporated into the resource-competition model. The parameters $c_{i\beta},m_{i},K_{\beta}$
are drawn independently for each value, and $\alpha_{ij}^{\left(d\right)}$
are drawn independently, except possibly a correlation between $\alpha_{ij}^{\left(d\right)}$
and $\alpha_{ji}^{\left(d\right)}$ controlling the symmetry of the
direct interactions. All quantities are drawn from Gaussian distributions,
parameterized by their first two moments. The definitions of parameters
are given in Appendix\textcolor{purple}{{} }\ref{sec:Basic-setup}.

As seen in Fig. \ref{fig:Summary-of-argument}(C), the variant with
even these small additional interactions shows persistent abundance
fluctuations, even if the ground model reaches equilibrium, Fig. \ref{fig:Summary-of-argument}(B).

\subsection{Theory for the onset of non-equilibrium dynamics\label{subsec:Theory-for-the}}

To understand whether and when the variant of the model will reach
a fixed point or a non-equilibrium state, we look at the stability
of fixed points, assuming they are reached. The basic idea is that
systems are sensitive to perturbations, if the fixed point in the
ground model is close to marginal stability.

Before turning to the present model, we review the results for the
random Lotka-Volterra models, which in the terminology of Sec. \ref{subsec:Sensitivity-to-direct}
only have ``direct'' interactions, $\alpha_{ij}=\alpha_{ij}^{\left(d\right)}$.
Their dynamics have been studied recently \citep{bunin_ecological_2017,biroli_marginally_2018,roy_numerical_2019,kessler_generalized_2015}
(see also related results in other models \citep{opper_phase_1992}).

A number of sharply delineated regions in parameter space are found,
referred to below as `phases'. In one phase the system reaches a unique
equilibrium. The boundary of this phase is marked by loss of stability
of these fixed points. Beyond this boundary (with a sharp transition
at large $S$) lies another phase, where the dynamics fail to reach
a fixed point and abundances fluctuate indefinitely \citep{roy_numerical_2019}.
In a special case where the interactions are symmetric, namely $\alpha_{ij}=\alpha_{ji}$,
this phase is instead characterized by with many possible alternative
equilibria, all of which are close to marginal stability \citep{biroli_marginally_2018}.

The behavior of the model variants defined above and in Sec. \ref{subsec:Resource-competition-with-non-li},
bares many similarities to that of the random Lotka-Volterra models.
There is a unique equilibrium phase, which is delineated by a boundary
at which the equilibrium looses its stability. Beyond it, we find
in simulations that the dynamics never reach a fixed point, as shown
in Figs. \ref{fig:Summary-of-argument},\ref{fig:Non-linear pert size}.
The case of symmetric $\alpha_{ij}^{\left(d\right)}$ is special and
appears to follow the scenario in \citep{biroli_marginally_2018},
see Appendix \ref{sec:Symmetric-additional-interaction}. An important
difference from random Lotka-Volterra models is the mechanism by which
fixed points loose their stability, which we now discuss.

We describe a method of calculating the species richness and stability
of the fixed points for the model variant described in Sec. \ref{subsec:Sensitivity-to-direct}.
This method is exact when the system admits a unique fixed point;
the loss of its stability marks the boundary of the phase. To highlight
the relation between marginal stability and sensitivity to perturbations,
we study the spectrum of the interaction matrix, and how it changes
for the model variant described above. A different approach, using
Dynamical Mean Field Theory, is possible and ultimately equivalent,
and has been employed on a related problem in \citep{yoshino_statistical_2007}.

Consider fixed points of the dynamics, i.e. abundance vectors $\vec{N}$
for which $d\vec{N}/dt=0$ in Eq. (\ref{eq:Reduced MCRM EOM}). We
are interested in the linear stability of these fixed points, namely
whether the system approaches the fixed point if initialized close
to it. The linear stability can be obtained from the properties of
the reduced interaction matrix $\alpha^{*}$ comprised only from interactions
between surviving species (for which $N_{i}\rightarrow c>0$, even
as the migration $\eta_{i}\rightarrow0$). A fixed point $\vec{N}$
is linearly stable if and only if all the real parts of eigenvalues
of $\alpha^{*}$ are positive, or equivalently if the minimal eigenvalue
real part is positive, $0<\min\left\{ \mathrm{Re}\left[\Lambda\left(\alpha^{*}\right)\right]\right\} \equiv\lambda_{min}$.
A fixed point is marginally stable if $\lambda_{min}\rightarrow0^{+}$
when $S\rightarrow\infty$.

While the MCRM only has stable or marginally stable fixed points,
$\lambda_{min}\ge0$, the model variant can have unstable ones. Close
to marginality, i.e. when $\lambda_{min}$ is zero or close to zero,
even a small perturbation may cause the system to lose its stability.
This is the case for a broad region in parameter space, as we now
show.

\begin{figure}
\begin{centering}
\includegraphics[width=1\columnwidth]{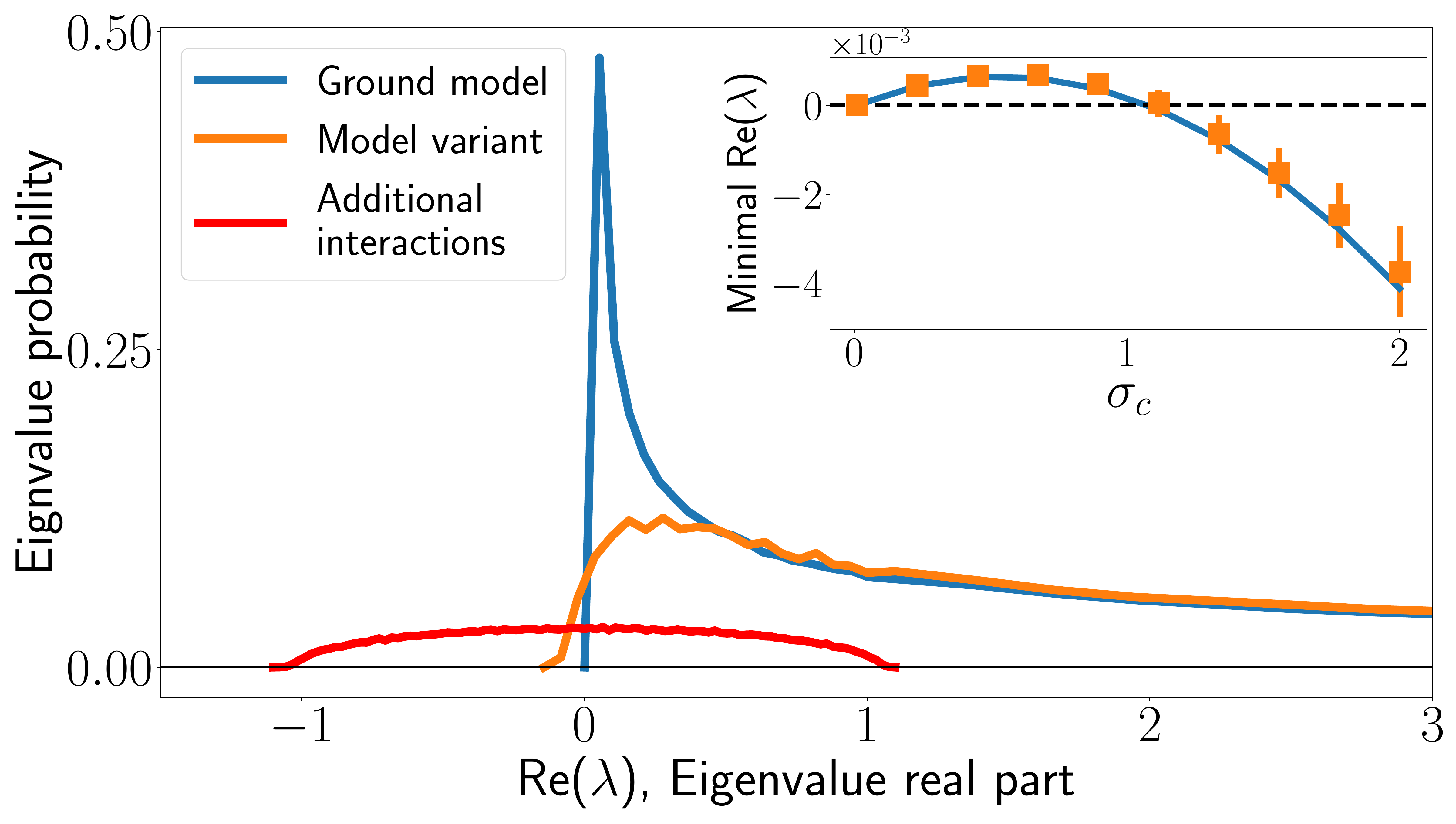}
\par\end{centering}
\caption{\label{fig:spectrum perturbation}Spectrum of $\alpha^{*}$, the interaction
matrix of persistent species, in the ground model (blue), and the
variant with additional direct interactions (orange). The perturbation
spectrum is shown in red, to illustrate its size we normalize the
area under the perturbation spectrum to the size of the relative perturbation
strength (0.05). (Inset) Minimal eigenvalue real part of the reduced
interaction matrix $\alpha^{*}$, when varying $\sigma_{c}$ at fixed
$\mu_{c}$. Solid line is theoretical curve. A phase transition occurs
when the minimal eigenvalue real part crosses from $\lambda_{min}>0$
at which fixed points of the dynamical system are stable, to $\lambda_{min}<0$
where all fixed points of the system are unstable, leading to persistent
dynamics.}
\end{figure}

\begin{figure*}
\begin{centering}
\includegraphics[width=1\textwidth]{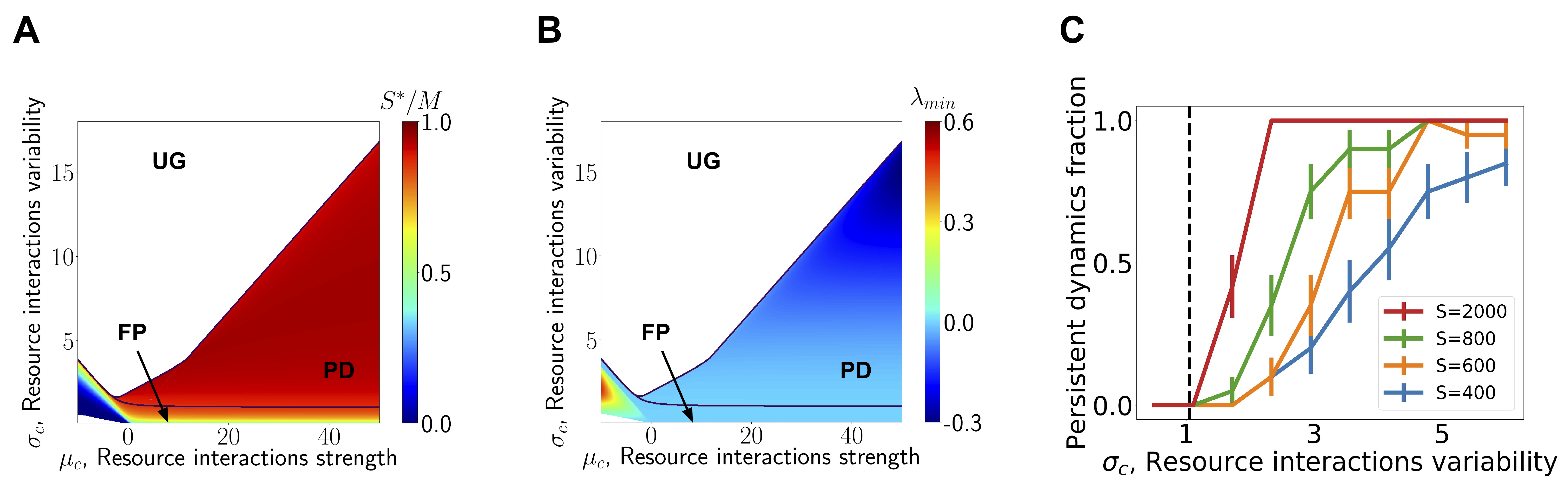}
\par\end{centering}
\caption{\label{fig:Phase diagram}The model exhibits three phases, i.e. regions
with qualitatively distinct behavior. In one, the system converges
to a stable fixed point (FP), in another fixed points of the system
are unstable yielding persistent dynamics (PD). In the third phase,
unbounded growth (UG), species abundances grow without bound. $\omega$
is adjusted in order to maintain constant perturbation strength of
$0.05$. (A) Color map of the ratio $S^{*}/M$, indicating how close
the system is to competitive exclusion $S^{*}/M=1$. (B) The minimal
real part of eigenvalues of the interaction matrix between coexisting
species, $\lambda_{min}$. Fixed point stability is lost at $\lambda_{min}=0$,
resulting in a phase transition to dynamically persistent states.
(C) Probability for having a persistent dynamics in a system with
interactions drawn for different values of $\sigma_{c}$, the variability
in consumer preferences. The transition between FP and PD phases becomes
sharper as system size increases.}
\end{figure*}

In Fig. \ref{fig:spectrum perturbation}, we show the spectrum of
such an $\alpha^{*}$ matrix close to a marginal fixed point. As expected,
the marginal case is characterized by non-vanishing density of eigenvalues
arbitrarily close to zero. When applying a small perturbation to the
marginal interaction matrix $\alpha^{*}$, for example $\alpha_{ij}^{\left(d\right)}$
described in Sec. \ref{subsec:Sensitivity-to-direct}, the spectrum
is broadened and may cross zero to give eigenvalues with negative
real parts, resulting in a dynamically unstable fixed point. The properties
of the fixed point depend crucially on the species richness (the number
of species that survive), which is a result of a balance between competition
that pushes species out of the system, and migration which allow them
to try and invade.%
\begin{comment}
This phenomena is robust due to the nature of generic perturbation
matrices which expected to have negative real eigenvalue parts.
\end{comment}

The method for calculating the spectrum consists of the following
main steps: first, we find the number $S^{*}$ of coexisting species
using the cavity method. This follows similar calculations in precious
works \citep{bunin_ecological_2017,advani_statistical_2018}, and
is detailed in Appendix \ref{sec:Cavity-equations}. We then calculate
$\lambda_{min}$, the minimal real part of the eigenvalues of $\alpha^{*}$,
for a reduced interaction matrix with $S^{*}$ species. This is done
using random matrix theory and detailed in Appendix \ref{sec:Random-Matrix-Theory}.%
\begin{comment}
It turns out that the marginality is determined by the system parameters
and $\phi$ alone {[}list required parameter combinations - $M/S,\mu_{c},\sigma_{c},\mu_{d},\sigma_{d},\gamma,K,\sigma_{K},m,\sigma_{m}${]}.
\end{comment}

Following this method, we can predict the dynamical behavior as a
function of the model parameters. We find three phases, shown in Fig.
\ref{fig:Phase diagram}.%
\begin{comment}
shows the phase diagram of the system at a cross section of $\sigma_{c}$
v.s $\mu_{c}$ with the other parameter held fixed.
\end{comment}
{} In the first, the system converges to a unique fixed point, independent
of the initial conditions, as in Fig. \ref{fig:Summary-of-argument}(B).
In the second, the system fails to reach a fixed point, with abundances
fluctuating indefinitely, as in Fig. \ref{fig:Summary-of-argument}(C).
In the third phase the abundances diverge, indicating that the model
is no longer adequate in this parameter regime.

Notably, when $S^{*}/M\approx1$ the unperturbed system is close to
competitive exclusion and correspondingly close to marginality, and
therefore the model variant with the direct interactions becomes unstable,
i.e. $\lambda_{min}<0$. As expected theoretically, the transition
between the two behaviors is sharp when $S,M$ are large, and happens
at the theoretically predicted value of the parameters, see Fig. \ref{fig:Phase diagram}(C).
In less diverse systems, the transition is more gradual.

The loss of stability of putative fixed points results in persistent
dynamics where species invade but are then pushed back out by the
instability of fixed points. This is clear in Fig. \ref{fig:Summary-of-argument}(C).

\subsection{Resource-competition with non-linear resource intake\label{subsec:Resource-competition-with-non-li}}

So far, we have discussed the ground model with a small addition of
other interactions. This allows us to identify regions in parameter
space where the ground model is sensitive to perturbations. By adding
interactions that are not mediated by resource competition, the model
variant can no longer be strictly interpreted as a resource-competition
model. Here we consider a second variant of the ground model, which
belongs to the resource-competition class, but with non-linear resource
intake. Non-linear dependence of the growth rate on resource availability
appears in many situations, such as in competition over essential
resources \citep{leon_competition_1975,huisman_oscillations_2002}.
Here the aim is not to study the consequences of a specific non-linear
mechanism, but rather to demonstrate the sensitivity of the model
to such variations in the dynamical rules.

We find that much like the model variant discussed in previous sections,
here too the dynamics are sensitive to the changes from the ground
model, with fixed points turning into persistent abundance fluctuations,
in much the same parameter regions as found previously.

\begin{figure}
\begin{centering}
\includegraphics[width=1\columnwidth]{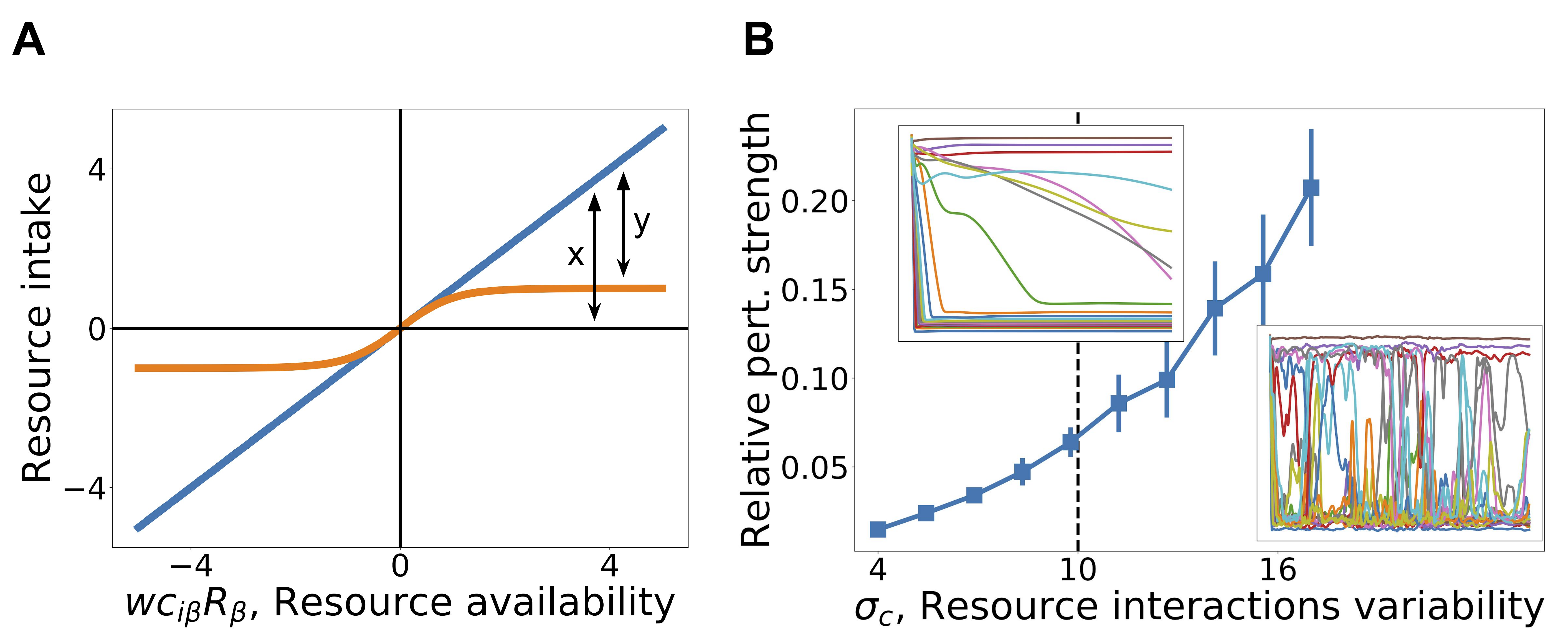}
\par\end{centering}
\caption{\label{fig:Non-linear pert size}A model with non-linear resource
intake. (A) Illustration of the non-linearity. The weighted linear
sum of the intake $wc_{i\beta}R_{\beta}$ of resource $\beta$ by
species $i$, used in the ground model, is replaced by a non-linear
function. The level of non-linearity, $p$, is measured by the ratio
$y/x$, see figure, where the $x$ axis denotes the intake and the
$y$-axis the linear- and non-linear intake functions, averaged over
species and time. (B) $p$ changes as model parameters are changed
(here varying $\sigma_{c}$). Persistent non-equilibrium dynamics
are found for $p\gtrsim0.06$.}
\end{figure}

The second variant to the ground model, Eq. (\ref{eq:MCRM EOM}),
is different from the ground model in the way that different resources
translate into the growth rate of the consumer. Whereas in Eq. (\ref{eq:MCRM EOM})
the growth rate is a linear combination of the resource values $R_{\beta}$,
here we use a non-linear function. We choose a non-linear consumption
function $h\left(R\right)=\frac{1}{w}\tanh\left(wR\right)$ with control
parameter $w$. With this consumption function, the dynamical equations
read:
\begin{equation}
\begin{cases}
\frac{dN_{i}}{dt} & =N_{i}\left[\sum_{\beta}\frac{1}{w}\tanh\left(wc_{i\beta}R_{\beta}\right)-m_{i}\right]+\eta_{i}\\
R_{\beta} & =K_{\beta}-\sum_{j}c_{j\beta}N_{j}
\end{cases}\label{eq:Tanh MCRM EOM}
\end{equation}
The parameter $w$ allows us to tune the deviation from the ground
model. For small values of $w$, $h\left(R\right)\simeq R$ so the
non-linear effects become small and the equations reduce to the ground
model Eq. (\ref{eq:MCRM EOM}). For finite $w$ non-linear effects
may be important. We quantify the deviation from linearity explored
by the dynamics by $p=1-\left\langle \frac{\tanh\left(wc_{i\beta}\bar{R_{\beta}}\right)}{wc_{i\beta}\bar{R_{\beta}}}\right\rangle _{i\beta}$,
where $\left\langle ..\right\rangle _{i\beta}$ is the average over
species and resources, and$\bar{R_{\beta}}$ denotes the time average
of the resource abundance $R_{\beta}$ (taken over a window of $\Delta t=1000$).
We find that even a rather small value of $p$ is sufficient to induce
a transition to dynamical persistence, see Fig. \ref{fig:Non-linear pert size},
where the transition occurs at $p\sim0.06$. Again, this demonstrates
how the system's dynamics may be sensitive to small changes in the
equations governing the model, in this case in a variant that is itself
strictly a resource competition model.

\subsection{Species abundance distribution and diversity\label{subsec:Intra-community-=000026-diversity}}

Above, we saw how systems near marginal stability are sensitive to
small variations in the model, either by additional interactions,
or by changes to the functional form of the interactions. Here we
show that these changes can allow the diversity to go well above the
number of resources. This is made possible by the persistent dynamics,
which are no longer bound by the competitive exclusion principle.

The competitive exclusion principle \citep{mcgehee_mathematical_1977}
states that for models describing an ecological community of $S$
species relying on $M$ limiting resources, no stable fixed points
with $M<S^{*}$ exist. Briefly, the core of the argument is that any
fixed point with $M<S^{*}$ would imply a degenerate Jacobian matrix
with rank $M$ or less. This kind of fixed point can be marginally
stable, but not stable. The second variant of the model, Eq. (\ref{eq:Tanh MCRM EOM}),
satisfies the conditions for this principle to hold, so the diversity
of stable equilibria is bound by $M$.

\begin{figure}
\begin{centering}
\includegraphics[width=1\columnwidth]{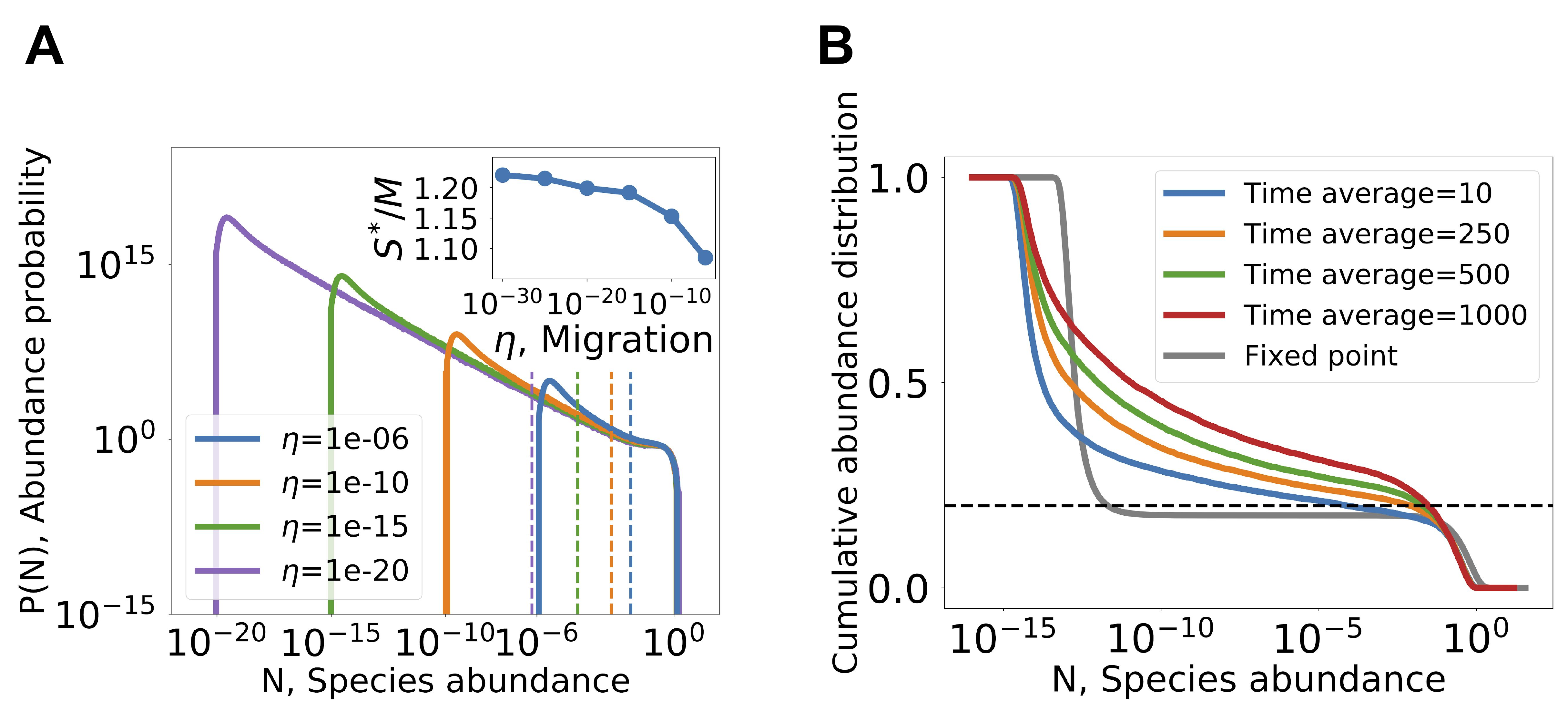}
\par\end{centering}
\caption{\label{fig:diversity-above-comp-ex}Diversity above competitive exclusion
in the non-linear RC model \ref{subsec:Resource-competition-with-non-li}.
(A) Abundance distribution for different values of migration. The
area to the right of the vertical lines hold exactly $M$ species.
The rest of the distribution, below the line, accounts for species
above competitive exclusion. Inset: the total number of coexisting
species normalized by number of resources, with values above one indicate
crossing of competitive exclusion. (B) Due to the abundance fluctuations,
averaging abundances over a time window pushes the distribution of
abundance upwards due to fluctuations. The cumulative abundance distribution
is shown, defined as $\mathcal{F}\left(N\right)=\frac{1}{S}\sum_{i}\ensuremath{1}_{\left[\bar{N}_{i}>N\right]}$.
The dashed line is the competitive exclusion bound, $M/S$. For comparison,
the distribution at a fixed point is given, showing that the number
of species at high abundance ($N\gtrsim10^{-3}$) does not reach this
bound, and the rest of the species are at low abundances, only supported
by migration.}
\end{figure}

As an example we look at the second model variant, as defined in Sec.
\ref{subsec:Resource-competition-with-non-li}. Long-time simulations
of the persistent dynamics show that the species abundance distribution
converges to a stationary form that can be decomposed into a power
law at intermediate abundances, and other parts at the highest and
lowest abundances, see Fig. \ref{fig:diversity-above-comp-ex}(A):

\begin{equation}
P\left(N\right)=\begin{cases}
P_{high}\left(N\right) & N_{u}<N\\
cN^{-\left(\nu+1\right)} & \eta\lesssim N<N_{u}\ .\\
P_{low}\left(N\right) & N\lesssim\eta
\end{cases}\label{eq:SAD}
\end{equation}
Here $N_{u}$ is a constant, and $c$ is set by the normalization
$\intop P\left(N\right)dN=1$. From the simulations, $\nu$ is not
far from zero when $\eta\rightarrow0$ ($\nu\approx0.02$ in Fig.
\ref{fig:diversity-above-comp-ex}, and similar for other parameter
sets, see Appendix \ref{sec:Simulation-details}).

We first ask about the instantaneous species richness, namely the
fraction of species that are not at the migration floor (say, above
$100\eta$). By integrating $P(N)$ in Eq. (\ref{eq:SAD}) one finds
that the fraction of the species above the migration floor approaches
a finite number when $\eta\rightarrow0^{+}$, for details see Appendix
\ref{sec:Species-abundance-distribution}. This number can be larger
than $M$, and in fact is so in the example shown in Fig. \ref{fig:diversity-above-comp-ex}.
In other words, a finite fraction of the species coexist above the
competitive exclusion limit even when migration is very small. This
is possible since the community is not in a fixed point, and so is
not bound by the competitive exclusion principle.

If the species abundance is measured by integrating over a finite-time
window, see Fig. \ref{fig:diversity-above-comp-ex}, the abundances
shift to higher values as the time window grows, indicating that species
have periods of time with high abundance. This leads to a growth in
the abundance $N_{CE}$ above which there are exactly $M$ species
with higher abundances.

In \citep{pearce_stabilization_2019}, chaotic dynamics where studied
in Lotka-Volterra equations with random interactions coefficients,
and the existence of of time periods with high abundance have been
reported, as well as a power law like in Eq. (\ref{eq:SAD}) (albeit
with a different exponent). The relation of these results to the present
resource-competition model are an interesting question for future
research.%
\begin{comment}
To show that, we compare the species diversity in the original MCRM
model, and in the two variants {[}or just one? what's in the figure?{]}{]}.
Diversity is quantified by two measures, the species richness (number
of species with $N_{i}>0$), \textcolor{red}{{[}should we introduce
here $\mathcal{F}\left(N\right)=\frac{1}{S}\sum_{i}\ensuremath{{1}}_{\left[\bar{N}_{i}>N\right]}$?
ref to fig{]}} and the Shannon index (...). When the original MCRM
system is far from marginal stability, the variants also reach a fixed
point. In this case there is little effect on diversity {[}{[}where
is this shown?{]}{]}. This demonstrates that the changes to the model
are indeed small. In contrast, when the MCRM is close to marginal
stability, the diversity of the variants is much higher than that
of the original model. Importantly, while the original model remains
below CE, the variants \textcolor{red}{{[}at the dynamical persistent
phase{]}} are able to greatly exceed it
\end{comment}

\section{Discussion and conclusions\label{subsec:Conclusions}}

In this work we have shown how diverse ecological communities with
resource-competition interactions may display non-equilibrium dynamics.
This turns out to be closely related to the ratio of realized species
diversity to the number of resources, $S^{*}/M$. When this number
is larger than one, fixed-points are either unstable or marginally
stable, as expected by the competitive exclusion principle. If they
are unstable, the system is pushed away from fixed points, and abundances
forever fluctuate. While marginal-stable fixed points are in principle
possible, they are structurally unstable under variations in the model,
such as non-linearities that destabilize the fixed points.

\subsection*{Comparison with random Lotka-Volterra models}

This picture bridges a gap to the behavior of high-dimensional models
where interactions are sampled at random without a specified mechanism.
In the notation of Sec. \ref{subsec:Sensitivity-to-direct}, this
corresponds to having $\alpha_{ij}=\alpha_{ij}^{\left(d\right)}$
only. These models show a phase with persistent dynamics \citep{opper_phase_1992,yoshino_statistical_2007,kessler_generalized_2015,bunin_ecological_2017,biroli_marginally_2018,roy_numerical_2019},
in contrast to resource-competition which have thus far only shown
relaxation to equilibrium in highly diverse communities \citep{tikhonov_collective_2017,advani_statistical_2018}.
We find that the generic phase-diagram is in fact much more similar,
with a transition to non-equilibrium dynamics when the variability
in interaction strengths is high enough (compare, for example, Fig.
\ref{fig:Phase diagram} with the phase diagram in \citep{bunin_ecological_2017}).%
\begin{comment}
{[}{[}rewrite?: Here we employ tools from statistical mechanics. Previous
relevant works: MacArthur models {[}Tikohonov \citep{tikhonov_collective_2017},
.., us \citep{advani_statistical_2018} - relation to competitive
exclusion \citep{landmann_systems_2018}, Pankaj -- below competitive
exclusion structure can increase stability \citep{cui_diverse_2019}.{]}.
Yoshino \citep{yoshino_statistical_2007} -- calculated the loss
of stability of the unique equilibrium, but did not make the connection
with the competitive exclusion principle, and did not explore the
region between that.{]}{]}
\end{comment}

One difference is that here, the symmetry of the interactions can
be very high and still lead to non-equilibrium dynamics. For example,
in the model with added direct interactions (Sec. \ref{subsec:Sensitivity-to-direct}),
the total interactions are very close to symmetric, with $\corr\left(\alpha_{ij},\alpha_{ji}\right)=0.997$.
In random Lotka-Volterra models, dynamics at a comparable level of
symmetry would typically relax to equilibria. This highlights the
importance of certain structures in the interaction network on dynamics.%
\begin{comment}
- If $\alpha_{ij}^{\left(d\right)}$ is also symmetric, then a different
scenario will occur, in which instead multiple fixed point will result
(returning to competitive exclusion??)
\end{comment}

\subsection*{Predictions}

How can the behavior discussed in this work be identified in natural
or experimental communities? The dynamical outcome will depend on
the following considerations:
\begin{itemize}
\item Is the community isolated; under migration from a regional species
pool; or part of a meta-community?
\item The ratio of realized species diversity to the number of resources
($S^{*}/M$).
\item Is the realized diversity $S^{*}$ high enough for high-diversity
effects to show?
\end{itemize}
\begin{comment}
\emph{experiments: are we seeing uninvadable solutions, or invadable
FPs? Can check by reintroducing initial inoculum. Two reasons why
this is not seen, even if we are in the right phase: small diversity
and single well-mixed community {[}{[}we still need the figure that
shows this{]}{]}.}
\end{comment}
Consider first a single well-mixed system with continuous migration
from a species pool, which was the focus of previous sections. In
such a setting, dynamics either a relax to single uninvadable equilibria
or reach persistent fluctuations. Which of these two possibilities
is realized depends on the system parameters: the realized species
diversity ($S^{*}$) is set by the balance between extinctions due
to competition and species able to invade. If fixed points at this
diversity are unstable, the latter outcome will result. This is the
Persistent Dynamics phase in Fig. \ref{fig:Phase diagram}. As shown
in Fig. \ref{fig:Phase diagram}, it is attained when there is sufficient
variability in the interactions, mediated for example by a broad distribution
of consumption preferences (high $\sigma_{c}$).%
\begin{comment}
The time required for a perturbed community to reach either of these
long-time behaviors is not known, and is an interesting problem for
future research.
\end{comment}
\begin{comment}
\begin{itemize}
\item an experiment can detect the persist fluctuations, given that enough
time is give for the system to relax to either a fixed point or to
persistent dynamics. The question of the required relaxation time
is not know, and is an interesting problem for future research.
\end{itemize}
\end{comment}

We turn to a single well-mixed community that is isolated (no migration,
$\eta_{i}=0$). Here species may go extinct due to large abundance
fluctuations, without being able to invade again. Extinctions may
then lead to equilibria even when non-equilibrium dynamics are expected
with migration, see Appendix \ref{sec:Meta-community}. The difference
is that these equilibria can be invaded by species from the species
pool. Importantly, in these conditions \emph{all} fixed points are
invadable, as uninvadable ones would translate to equilibria in the
presence of migration. If there are now isolated migration events
from the species pool that are well-separated in time (for example,
at low migration rates, or in experiments where species are re-introduced)
the equilibria will be punctuated by migration events that change
the community composition \citep{law_alternative_1993}.%
\begin{comment}
\begin{itemize}
\item Then, species may be go extinct in the community. If sufficiently
many species go extinct, the system may then reach a fixed point.
The question then is whether the fixed points are invadable: whether
species from the original pool may grow if they invade at small abundances.
In the Unique Equilibrium phase there is a single uninvadable fixed
point, that can be reached if all species attempt to invade and the
population sizes are large enough. In the Persistent Dynamics phase,
no such fixed point exists: all fixed points can be invaded by at
least one species. Thus, if one injects species from the pool at discrete
times, the species composition will continue to change in a ``meta-dynamics''
between such injections {[}{[}ref: morton\_regional\_1997 ?{]}{]}.
\end{itemize}
\end{comment}
\begin{comment}
meta-community: refer to appendix with persistent solutions {[}{[}this
can follow from the previous point{]}{]}.
\end{comment}

An explicit spatial dimension, such as a meta-community in which several
well-mixed systems are coupled by migration, again changes the phenomenology.
In this case, one might also find persistent fluctuations for a meta-community,
even if it is isolated from any outside species pool, allowing species
to go extinct within it. Still, the remaining species might continue
to fluctuate for extremely long times without inducing extinctions.
This has been shown recently for many-species meta-communities with
random Lotka-Volterra interactions in \citep{roy_can_2019,pearce_stabilization_2019}.
An example simulation, provided as a proof-of-principle, is provided
in Appendix \ref{sec:Meta-community}. The conditions for non-equilibrium
dynamics to persist depend on additional parameters including the
migration rates and the number of communities in the meta-community.
A fuller account of this effect is an interesting direction for future
research.

Finally, we note that the non-equilibrium dynamics discussed in this
work apply to communities with many species and resources or niches.
Simulations indicate that dynamical fluctuations appear when there
are tens of species in the community or more; communities with fewer
species may instead relax to equilibria.

High-dimensional ecological dynamics are, in some respects, qualitatively
different from their low-dimensional counterparts. Here we classified
possible scenarios for the dynamics of resource-competition communities,
and provided predictions for each scenario. We hope it may help in
guiding future theoretical works, observations and experiments on
high-diversity communities.

\emph{Acknowledgments} - It is a pleasure to thank J.-F. Arnoldi,
M. Barbier and G. Biroli for helpful discussions. G. Bunin acknowledges
support by the Israel Science Foundation (ISF) Grant no. 773/18.

\bibliographystyle{unsrt}
\bibliography{refs_from_Guy}
\clearpage{}

\appendix
%dummy comment inserted by tex2lyx to ensure that this paragraph is not empty%dummy comment inserted by tex2lyx to ensure that this paragraph is not empty

\section{Basic setup\label{sec:Basic-setup}}

Each instance of the first model variant, defined in Sec. \ref{subsec:Sensitivity-to-direct},
requires setting the values of the quantities $\left\{ c_{i\beta},\alpha_{ij}^{\left(d\right)},m_{i},K_{\beta}\right\} $,
referred to here as the disorder parameters. They are fixed from the
control parameters $\left\{ S,M,\mu_{c},\sigma_{c},\mu_{d},\sigma_{d},\gamma,\mu_{m},\sigma_{m},\mu_{K},\sigma_{K}\right\} $
as follows.

Let us denote by $\left\langle X\right\rangle $ the expectation value
of random variable $X$. All results cited in the paper at high-diversity
depend only on the first and second moment of the system disorder
parameters distribution. The means and variances of these are given
by $\left\langle c_{i\beta}\right\rangle =\mu_{c}/S,\ \left\langle \left(c_{i\beta}-\left\langle c_{i\beta}\right\rangle \right)^{2}\right\rangle =\sigma_{c}^{2}/S,\ \left\langle m_{i}\right\rangle =m,\ \left\langle \left(m_{i}-\left\langle m_{i}\right\rangle \right)^{2}\right\rangle =\sigma_{m}^{2},\ \left\langle K_{\beta}\right\rangle =K,\ \left\langle \left(K_{\beta}-\left\langle K_{\beta}\right\rangle \right)^{2}\right\rangle =\sigma_{K}^{2}$,
and the direct interactions by $\left\langle \alpha_{ij}^{\left(d\right)}\right\rangle =\mu_{d}/S,\ \left\langle \left(\alpha_{ij}^{\left(d\right)}-\left\langle \alpha_{ij}^{\left(d\right)}\right\rangle \right)^{2}\right\rangle =\sigma_{d}^{2}/S$
and $\corr\left(\alpha_{ij}^{\left(d\right)},\alpha_{ji}^{\left(d\right)}\right)=\gamma$
with $-1\leq\gamma\leq1$. All other cumulants are set to zero. This
definition of the parameters ensures that the abundance distribution
$P\left(N_{i}\right)$ and the fraction of persistent species have
a finite, well-defined limit as $S,M$ are taken to be large. In other
words, in that limit all results will only depend on these control
parameter combinations, e.g. on $\mu_{c}=S\left\langle c_{i\beta}\right\rangle $
rather than on $S,\left\langle c_{i\beta}\right\rangle $ separately.
The same results are obtained for a sparse interaction matrix with
$C$ non-zero links per species, as long as $1\ll C$. In that case,
which includes the case $C=S$ above, the moments are rescaled by
$C$ rather than $S$, e.g. $\dot{\left\langle c_{i\beta}\right\rangle =\mu_{c}/C}$
instead of $\left\langle c_{i\beta}\right\rangle =\mu_{c}/S$.

To simplify the notation below, it is useful to separate quantities
into mean and fluctuating parts
\[
c_{i\beta}\equiv\frac{\mu_{c}}{S}+\sigma_{c}d_{i\beta}\ ;\enskip\left\langle d_{i\beta}\right\rangle =0\enskip;\ \left\langle d_{i\alpha}d_{j\beta}\right\rangle =\frac{\delta_{ij}\delta_{\alpha\beta}}{S}
\]

\[
\alpha_{ij}^{\left(d\right)}=\frac{\mu_{d}}{S}+\sigma_{d}a_{ij}\enskip;\ \left\langle a_{ij}\right\rangle =0\enskip;\ \left\langle a_{ij}^{2}\right\rangle =\frac{1}{S}\enskip;\ \left\langle a_{ij}a_{ji}\right\rangle =\frac{\gamma}{S}
\]
\[
K_{\beta}=K+\delta K_{\beta}\enskip;\ \left\langle K_{\beta}\right\rangle =K\enskip;\ \left\langle \delta K_{\alpha}\delta K_{\beta}\right\rangle =\delta_{\alpha\beta}\sigma_{K}^{2}
\]
\[
m_{i}=m+\delta m_{i}\enskip;\ \left\langle m_{i}\right\rangle =m\enskip;\ \left\langle \delta m_{i}\delta m_{j}\right\rangle =\delta_{ij}\sigma_{m}^{2}
\]
With these definitions the first variant, Eq. (\ref{eq:Reduced MCRM EOM}),
can be written as

\[
\begin{cases}
\frac{dN_{i}}{dt}=N_{i}\left[g+\sigma_{c}\sum_{\beta}d_{i\beta}R_{\beta}-\omega\sigma_{d}\sum_{j}a_{ij}^{\left(d\right)}N_{j}-\delta m_{i}\right]+\eta_{i}\\
R_{\beta}=K^{eff}+\delta K_{\beta}-\sigma_{c}\sum_{j}d_{j\beta}N_{j}
\end{cases}
\]
where
\[
\left\langle R\right\rangle =\frac{1}{M}\sum_{\alpha}R_{\alpha}\quad\left\langle N\right\rangle =\frac{1}{S}\sum_{j}N_{j}
\]
\[
g=\mu_{c}\frac{M}{S}\left\langle R\right\rangle -\omega\mu_{d}\left\langle N\right\rangle -m
\]
\[
K^{eff}=K-\mu_{c}\left\langle N\right\rangle 
\]
This form of the equations will be useful below, in Appendix \ref{sec:Cavity-equations}.

\section{Model definition, parameters and simulation details\label{sec:Simulation-details}}

Differential equations were integrated using a Radau integrator implemented
in Python's Scipy package. Absolute integration tolerance is set to
$atol=0.1\eta$, where $\eta$ is the migration strength. Initial
conditions of species abundances are drawn from uniform distribution
over $\left[0,1\right]$. Perturbation strength is controlled using
$\omega$ to satisfy $\left\Vert \omega\cdot\alpha^{\left(d\right)}\right\Vert _{F}/\left\Vert \alpha^{\left(r\right)}\right\Vert _{F}=0.05$
throughout. Simulation parameters are summarized in Table \ref{tab:simulation_parameters}.

A Python code that run simulations of the ground model and its two
variants, with example parameters for the Fixed Point and Persistent
Dynamics phases, is given in: \href{https://github.com/Itaydal/crm-chaos}{https://github.com/Itaydal/crm-chaos}.

\onecolumngrid

\begin{table}[H]
\centering{}%
\begin{tabular}{|c|c|c|c|c|c|c|c|c|c|c|c|}
\hline 
 & $S$ & $M$ & $S/M$ & $\mu_{c}$ & $\sigma_{c}$ & $\mu_{d}$ & $\sigma_{d}$ & $K$ & $m$ & $\eta$ & $w$ (Sec. \ref{subsec:Resource-competition-with-non-li})\tabularnewline
\hline 
\hline 
Fig. \ref{fig:Summary-of-argument} (B) & 800 & 160 & 5 & 20 & 0.5 & 10 & 20 & 2 & 0.2 & $10^{-15}$ & -\tabularnewline
\hline 
Fig. \ref{fig:Summary-of-argument} (C) & 800 & 160 & 5 & 20 & 4 & 10 & 20 & 2 & 0.2 & $10^{-15}$ & -\tabularnewline
\hline 
Fig. \ref{fig:spectrum perturbation} & 1000 & 200 & 5 & 30 & 10 & 10 & 20 & 2 & 0.2 & - & -\tabularnewline
\hline 
Fig. \ref{fig:Phase diagram} (A,B) & $\infty$ & $\infty$ & 5 & Varying & Varying & 10 & 20 & 2 & 0.2 & - & -\tabularnewline
\hline 
Fig. \ref{fig:Phase diagram} (C) & Varying & Varying & 5 & 20 & Varying & 10 & 20 & 2 & 0.2 & $10^{-15}$ & -\tabularnewline
\hline 
Fig. \ref{fig:Non-linear pert size} & 800 & 160 & 5 & 30 & Varying & - & - & 2 & 0.2 & $10^{-13}$ & $0.05\cdot S$\tabularnewline
\hline 
Fig. \ref{fig:diversity-above-comp-ex} & 800 & 160 & 5 & 50 & 20 & - & - & 5 & 0.2 & Varying & $0.05\cdot S$\tabularnewline
\hline 
Fig. \ref{fig:s_star_vs_cavity} & 800 & 160 & 5 & 20 & 4 & 10 & 20 & 2 & 0.2 & $10^{-15}$ & -\tabularnewline
\hline 
Fig. \ref{fig:fluctuations_vs_directions} & 800 & 56 & 14.3 & 20 & 4 & 10 & 20 & 2 & 0.2 & $10^{-15}$ & -\tabularnewline
\hline 
Fig. \ref{fig:dynamics-no-migration} & 800 & 160 & 5 & 20 & 4 & 10 & 20 & 2 & 0.2 & $0$ & -\tabularnewline
\hline 
Fig. \ref{fig:min_eigval_gamma_one-fit} & Varying & Varying & 2 & 40 & 23 & 10 & 10 & 3 & 5 & $10^{-15}$ & -\tabularnewline
\hline 
Fig. \ref{fig:min_eigval_gamma_one} & $\infty$ & $\infty$ & 2 & 40 & Varying & 10 & 10 & 3 & 5 & $10^{-15}$ & -\tabularnewline
\hline 
\end{tabular}\caption{\label{tab:simulation_parameters} Simulation parameters used to create
each of the figures. The parameters $\gamma,\sigma_{K},\sigma_{m}$
are set to zero throughout.}
\end{table}

\twocolumngrid

\section{Cavity equations \label{sec:Cavity-equations}}

To study the properties of a typical fixed point of the model, we
use a variant of the cavity method \citep{rieger_solvable_1989,opper_phase_1992,mezard_sk_1986,crisanti_sphericalp-spin_1993,bunin_ecological_2017,advani_statistical_2018}.
It proceeds by adding one new species and one new resource, along
with newly sampled interactions between it and the rest of the system,
creating an $S+1$ species system with $M+1$ resources. Then, by
comparing the properties of a typical species of the old system with
those of the newly added species and resource we get self-consistent
equations for the macroscopic variables $\phi,\left\langle N\right\rangle ,\left\langle N^{2}\right\rangle $
where $\phi=S^{*}/S$ is the fraction of living species, together
with the properties of the resources.

Solving these self-consistent Eq. (\ref{eq:self_consistency}) for
range of parameters allows us to derive the phase diagram in Fig.
\ref{fig:Phase diagram}. In particular, the distinction between stable
and non-equilibrium phases is done by solving for $\phi$ for some
choice of control parameters, this determines the distribution of
reduced interaction matrices, in Appendix \ref{sec:Random-Matrix-Theory}
we calculate it's stability. The transition into the unbounded growth
phase is found at the divergence of $\left\langle N\right\rangle $. 

\subsection{Deriving species and resource distributions using cavity method}

Introducing to the system new resource and species $R_{0}$ and $N_{0}$
\begin{align*}
\frac{1}{N_{i}}\frac{dN_{i}}{dt}= & \left[g+\sigma_{c}\sum_{\alpha}d_{i\alpha}R_{\alpha}+\sigma_{c}d_{i0}R_{0}\right.\\
 & \left.-\delta m_{i}-\omega\sigma_{d}\sum_{j}a_{ij}N_{j}-\omega\sigma_{d}a_{i0}N_{0}\right]\\
R_{\alpha}= & K^{eff}-\sigma_{c}\sum_{j}d_{j\alpha}N_{j}+\delta K_{\alpha}-\sigma_{c}d_{0\alpha}N_{0}
\end{align*}
and the corresponding equations for $R_{0}$ and $N_{0}$ are

\begin{align*}
\frac{1}{N_{0}}\frac{dN_{0}}{dt}= & \left[g+\sigma_{c}\sum_{\alpha}d_{0\alpha}R_{\alpha}+\sigma_{c}d_{00}R_{0}\right.\\
 & \left.-\delta m_{0}-\omega\sigma_{d}\sum_{j}a_{0j}N_{j}-\omega\sigma_{d}a_{00}N_{0}\right]\\
R_{0}= & K^{eff}-\sigma_{c}\sum_{j}d_{j0}N_{j}+\delta K_{0}-\sigma_{c}d_{00}N_{0}
\end{align*}
Denote the steady-state value of a quantity $X$ by $\bar{X}$, also
denote by $\bar{X}_{\backslash0}$ the steady-state value of $X$
in the absence of the resource and species $'0'$.

Then we can define the following susceptibilities 
\begin{align*}
\chi_{i\beta}^{\left(N\right)}=\frac{\partial\bar{N}_{i}}{\partial K_{\beta}} & \:;\qquad\chi_{\alpha\beta}^{\left(R\right)}=\frac{\partial\bar{R}_{\alpha}}{\partial K_{\beta}}\\
\nu_{ij}^{\left(N\right)}=\frac{\partial\bar{N}_{i}}{\partial m_{j}} & \negmedspace;\qquad\nu_{\alpha j}^{\left(R\right)}=\frac{\partial\bar{R}_{\alpha}}{\partial m_{j}}
\end{align*}
Since addition of single resource and species is a small (order $S^{-1}$)
perturbation we can write

\begin{align*}
\bar{N}_{i}= & \left[\bar{N}_{i\backslash0}-\sigma_{c}\sum_{\beta}\chi_{i\beta}^{\left(N\right)}d_{0\beta}N_{0}\right.\\
 & \left.-\sum_{j}\nu_{ij}^{\left(N\right)}\left(\sigma_{c}d_{j0}R_{0}-\omega\sigma_{d}a_{j0}N_{0}\right)\right]\\
\bar{R}_{\alpha}= & \left[\bar{R}_{\alpha\backslash0}-\sigma_{c}\sum_{\beta}\chi_{\alpha\beta}^{\left(R\right)}d_{0\beta}N_{0}\right.\\
 & \left.-\sum_{j}\nu_{\alpha j}^{\left(R\right)}\left(\sigma_{c}d_{j0}R_{0}-\omega\sigma_{d}a_{j0}N_{0}\right)\right]
\end{align*}
We can now plug in these expressions into the steady-state equations
for $N_{0}$ and $R_{0}$. By taking leading order contributions to
$S^{-1}$, and take expectation value over expressions we get

\begin{align*}
0 & =\bar{N}_{0}\left[g-\frac{\sigma_{c}^{2}}{S}\sum_{\alpha}\chi_{\alpha\alpha}^{\left(R\right)}N_{0}-\omega^{2}\sigma_{d}^{2}\frac{\gamma}{S}\sum_{j}\nu_{jj}^{\left(N\right)}N_{0}\right.\\
 & \left.+\sigma_{c}\sum_{\alpha}d_{0\alpha}\bar{R}_{\alpha\backslash0}-\omega\sigma_{d}\sum_{j}a_{0j}\bar{N}_{j\backslash0}-\delta m_{0}\right]
\end{align*}
Notice that, to leading order in $S^{-1}$, as sum of weakly interacting
terms we can model the expression $\sigma_{c}\sum_{\alpha}d_{0\alpha}\bar{R}_{\alpha\backslash0}-\omega\sigma_{d}\sum_{j}a_{0j}\bar{N}_{j\backslash0}-\delta m_{0}$
as a Gaussian random field with mean $0$ and variance
\[
\sigma_{g}^{2}=\sigma_{c}^{2}\frac{M}{S}q_{R}+\omega^{2}\sigma_{d}^{2}q_{N}+\sigma_{m}^{2}
\]
where
\[
q_{N}=\frac{1}{S}\sum_{j}\bar{N}_{j\backslash0}^{2}\qquad q_{R}=\frac{1}{M}\sum_{\alpha}\bar{R}_{\alpha\backslash0}^{2}
\]
Let $z_{N}$ be a Gaussian random field with zero mean and unit variance,
and define the average susceptibilities

\[
\chi=\frac{1}{M}\sum_{\alpha}\chi_{\alpha\alpha}^{\left(R\right)}\qquad\nu=\frac{1}{S}\sum_{j}\nu_{jj}^{\left(N\right)}
\]
As there is no difference between species `0' and the rest, we can
emit the subscript '0' to and write the equation the fixed point abundance
distribution
\[
0=\bar{N}\left[g-\left(\sigma_{c}^{2}\frac{M}{S}\chi+\omega^{2}\sigma_{d}^{2}\gamma\nu\right)\bar{N}+\sigma_{g}z_{N}\right]
\]

Following similar procedure for the resources yields 

\[
0=\bar{R}\left[K^{eff}-\left(1-\sigma_{c}^{2}\nu\right)\bar{R}+\sigma_{K^{eff}}z_{R}\right]
\]
\[
\sigma_{K^{eff}}^{2}=\sigma_{K}^{2}+\sigma_{c}^{2}q_{N}
\]
We can solve these equations and get
\[
\bar{N}=\frac{\max\left[0,g+\sigma_{g}z_{N}\right]}{\frac{M}{S}\sigma_{c}^{2}\chi+\omega^{2}\gamma\sigma_{d}^{2}\nu}
\]
\[
\bar{R}=\frac{K^{eff}+\sigma_{K^{eff}}z_{R}}{1-\sigma_{c}^{2}\nu}
\]

\subsection{Self consistent equations}

At this stage, our aim is to solve for $\left\{ \phi_{S},\left\langle N\right\rangle ,\left\langle R\right\rangle ,q_{N},q_{R},\chi,\nu\right\} $
for a given set of control parameters $\left\{ S,M,K,\sigma_{K},m,\sigma_{m},\mu_{c},\sigma_{c},\mu_{d},\sigma_{d},\gamma,\omega\right\} $.
To that end, it is helpful to define
\[
\Delta_{g}=\frac{g}{\sigma_{g}}=\frac{\mu_{c}\frac{M}{S}\left\langle R\right\rangle -\omega\mu_{d}\left\langle N\right\rangle -m}{\sqrt{\sigma_{c}^{2}\frac{M}{S}q_{R}+\omega^{2}\sigma_{d}^{2}q_{N}+\sigma_{m}^{2}}}
\]
and the function
\[
w_{j}\left(\Delta\right)=\intop_{-\Delta}^{\infty}\frac{dz}{\sqrt{2\pi}}e^{-\frac{z^{2}}{2}}\left(z+\Delta\right)^{j}
\]
note that for $y=\max\left[0,\frac{a+c\cdot z}{b}\right]$ with $z$
Gaussian random variable we have that
\[
\left\langle y^{j}\right\rangle =\left(\frac{c}{b}\right)^{j}\intop_{-\frac{a}{c}}^{\infty}\frac{dz}{\sqrt{2\pi}}e^{-\frac{z^{2}}{2}}\left(z+\frac{a}{c}\right)^{j}=\left(\frac{c}{b}\right)^{j}w_{j}\left(\frac{a}{c}\right)
\]
Taking the first two moments of the distributions $\bar{N}$ and $\bar{R}$,
leads to the set of set consistent equations
\begin{align}
\phi_{S} & =w_{0}\left(\Delta_{g}\right)\nonumber \\
\left\langle N\right\rangle  & =\left(\frac{\sigma_{g}}{\frac{M}{S}\sigma_{c}^{2}\chi+\omega^{2}\gamma\sigma_{d}^{2}\nu}\right)w_{1}\left(\Delta_{g}\right)\nonumber \\
\left\langle R\right\rangle  & =\frac{K^{eff}}{1-\sigma_{c}^{2}\nu}=\chi K^{eff}\nonumber \\
q_{N} & =\left\langle N^{2}\right\rangle =\left(\frac{\sigma_{g}\nu}{\frac{M}{S}\sigma_{c}^{2}\chi+\omega^{2}\gamma\sigma_{d}^{2}\nu}\right)^{2}w_{2}\left(\Delta_{g}\right)\label{eq:self_consistency}\\
q_{R} & =\left\langle R^{2}\right\rangle =\chi^{2}\left[\sigma_{K^{eff}}^{2}+\left(K^{eff}\right)^{2}\right]\nonumber \\
\nu & =\left\langle \frac{\partial\bar{N}}{\partial m}\right\rangle =-\frac{\phi_{S}}{\frac{M}{S}\sigma_{c}^{2}\chi+\omega^{2}\gamma\sigma_{d}^{2}\nu}\nonumber \\
\chi & =\left\langle \frac{\partial\bar{R}}{\partial K}\right\rangle =\frac{1}{1-\sigma_{c}^{2}\nu}\nonumber 
\end{align}
The expressions for $\nu$ and $\chi$ are derived by differentiating
abundances distributions $N,R$ with respect to $m$ and $K$ respectively
and taking their expectation values. 

To avoid singularities at the diverging phase ($\left\langle N\right\rangle \rightarrow\infty$)
we define $h=\frac{1}{\left\langle N\right\rangle },\ q_{n}=\frac{\left\langle N^{2}\right\rangle }{\left\langle N\right\rangle ^{2}}=\frac{q_{N}}{\left\langle N\right\rangle ^{2}}$.
With these variables the self consistent equations read

\begin{align*}
\phi_{S} & =w_{0}\left(\Delta_{g}\right)\\
h & =-\frac{1}{\nu\sigma_{g}}\frac{w_{0}\left(\Delta_{g}\right)}{w_{1}\left(\Delta_{g}\right)}\\
\left\langle R\right\rangle  & =\frac{1}{h}\chi\left(Kh-\mu_{c}\right)\\
q_{n} & =h^{2}\left(\frac{\sigma_{g}\nu}{\phi_{S}}\right)^{2}w_{2}\left(\Delta_{g}\right)=\frac{w_{2}\left(\Delta_{g}\right)}{\left[w_{1}\left(\Delta_{g}\right)\right]^{2}}\\
q_{R} & =\frac{1}{h^{2}}\chi^{2}\left[\left(K^{2}+\sigma_{K}^{2}\right)h^{2}-2\mu_{c}Kh+\sigma_{c}^{2}q_{n}+\mu_{c}^{2}\right]\\
\nu & =-\frac{\phi_{S}}{\frac{M}{S}\sigma_{c}^{2}\chi+\omega^{2}\gamma\sigma_{d}^{2}\nu}\\
\chi & =\frac{1}{1-\sigma_{c}^{2}\nu}
\end{align*}

At this stage one has to find a self consistent solution for this
set of equations. One possible approach would be to use a global numerical
optimizer such as a basin-hopping algorithm to find a solution in
the 7-dimensional parameter space spanned by $\left\{ \phi_{S},h,\left\langle R\right\rangle ,q_{n},q_{R},\nu,\chi\right\} $.
This requires non-convex optimization in high dimension, which is
not guaranteed to work. By some additional manipulation we were able
to reduce it into a one dimensional non-convex optimization over the
variable $\Delta_{g}$, as we now show.

Simplifying the expressions for the susceptibilities results with
the third order polynomial for $\nu$ where the only unknown is $\phi_{S}$. 

\[
\omega^{2}\gamma\sigma_{d}^{2}\sigma_{c}^{2}\nu^{3}-\omega^{2}\gamma\sigma_{d}^{2}\nu^{2}-\left(\frac{M}{S}\sigma_{c}^{2}-\phi_{S}\sigma_{c}^{2}\right)\nu-\phi_{S}=0
\]
Note that $\phi_{S}$ only depends on $\Delta_{g}$, therefore one
can span a grid of values for $\Delta_{g}$ and assigning the roots
the above polynomial for each $\nu_{i}\left(\Delta_{g}\right)$ where
$i=1,2,3$. Plugging back into the expression for resources susceptibility
leads to $\chi_{i}\left(\Delta_{g}\right)$.

Now, using the relations $\sigma_{g}h=-\frac{1}{\nu}w_{0}\left(\Delta_{g}\right)/w_{1}\left(\Delta_{g}\right)$
and $\Delta_{g}=\frac{g}{\sigma_{g}}$ yields
\[
\left(\frac{M}{S}\mu_{c}\chi K-m\right)h-\frac{M}{S}\mu_{c}^{2}\chi-\omega\mu_{d}=-\Delta_{g}\frac{1}{\nu}\frac{w_{0}\left(\Delta_{g}\right)}{w_{1}\left(\Delta_{g}\right)}
\]
solving this for $h\left(\Delta_{g}\right)$ leads to 
\[
h_{i}\left(\Delta_{g}\right)=\frac{\frac{M}{S}\mu_{c}^{2}\chi_{i}\left(\Delta_{g}\right)+\omega\mu_{d}-\frac{\Delta_{g}}{\nu_{i}\left(\Delta_{g}\right)}\frac{w_{0}\left(\Delta_{g}\right)}{w_{1}\left(\Delta_{g}\right)}}{\frac{M}{S}\mu_{c}K\chi_{i}\left(\Delta_{g}\right)-m}
\]
Rewriting the expression for $\Delta_{g}$ with the new variables
$h,q_{n}$ 

\[
g=\frac{1}{h}\left[\frac{M}{S}\mu_{c}\chi\left(Kh-\mu_{c}\right)-\omega\mu_{d}-mh\right]
\]
\[
\sigma_{g}=\frac{1}{h}\left\{ \begin{array}{c}
\frac{M}{S}\chi^{2}\sigma_{c}^{2}\left[\sigma_{K}^{2}h^{2}+\sigma_{c}^{2}q_{n}+\left(Kh-\mu_{c}\right)^{2}\right]\\
+\sigma_{m}^{2}h^{2}+\omega^{2}\sigma_{d}^{2}q_{n}
\end{array}\right\} ^{1/2}
\]

\begin{eqnarray*}
\hat{\Delta}_{g} & = & \frac{\frac{M}{S}\mu_{c}\chi\left(Kh-\mu_{c}\right)-\omega\mu_{d}-mh}{\left\{ \begin{array}{c}
\frac{M}{S}\chi^{2}\sigma_{c}^{2}\left[\sigma_{K}^{2}h^{2}+\sigma_{c}^{2}q_{n}+\left(Kh-\mu_{c}\right)^{2}\right]\\
+\sigma_{m}^{2}h^{2}+\omega^{2}\sigma_{d}^{2}q_{n}
\end{array}\right\} ^{1/2}}
\end{eqnarray*}
Finally, find values of $\Delta_{g}$ and $i=1,2,3$ where $\hat{\Delta}_{g}\left[h_{i}\left(\Delta_{g}\right)\right]=\Delta_{g}$.
With these self consistent values for $\Delta_{g},h,\nu,\chi$ it
is straight forward to then find $q_{n},\left\langle R\right\rangle ,q_{R}$.

\section{Random Matrix Theory \label{sec:Random-Matrix-Theory}}

Given the values of control parameters as described in Appendix \ref{sec:Basic-setup},
the diversity $\phi=S^{*}/S$ for the perturbed MCRM (Sec. \ref{subsec:Sensitivity-to-direct})
can be found as described in Appendix \ref{sec:Cavity-equations}.
Here we define the random matrix ensemble corresponding to the reduced
interaction matrix for given control parameters and diversity values.
The main result of this appendix is the minimal eigenvalue real part
of the ensemble Eq. (\ref{eq:eigenvalue_contour_real}) in Appendix
\ref{subsec:Calculating-minimal-eigval}. This in turn is used to
distinguish between the stable and non-equilibrium phases in Fig.
\ref{fig:Phase diagram}. 

\subsection{Random matrix theory and free probability}

The linear stability of a fixed point is determined by the sign of
the minimal eigenvalue of its interaction matrix. For randomly sampled
interaction matrices, the problem of determining the sign of the minimal
eigenvalue can be addressed with random matrix theory (RMT). A random
matrix is a matrix whose elements are drawn from probability distribution,
known as an ensemble. One of the main uses of RMT is to determine
what the spectrum of a typical matrix drawn from such ensemble would
look like, and in particular its minimal eigenvalue. Below we describe
the key steps taken to find the minimal eigenvalue of the particular
ensemble at hand. For a detailed review of these techniques see \citep{livan_introduction_2018}.

A central object in RMT is the Green function of an ensemble, also
known as a Resolvent or Stieltjes transform. For an $N\times N$ random
matrix $H$, the Green function is defined as
\[
G_{N}\left(z\right)=\frac{1}{N}Tr\left(\left[z\mathbb{I}-H\right]^{-1}\right)=\frac{1}{N}\sum_{i=1}^{N}\frac{1}{z-x_{i}}
\]
where $x_{1},\ldots x_{N}$ are the eigenvalues of $H$. Since $H$
is a random matrix, $G_{N}\left(z\right)$ is a random complex function
with poles at locations $x_{i}$. There are several methods for deriving
the green function for a given ensemble, for details see \citep{livan_introduction_2018}.
Averaging over $H$ and taking the thermodynamic limit ($N\rightarrow\infty$),
\[
G\left(z\right)=\lim_{N\rightarrow\infty}\left\langle G_{N}\left(z\right)\right\rangle =\intop dx\frac{\rho\left(x\right)}{z-x}\ .
\]
At the thermodynamic limit the set of eigenvalues $x_{1},\ldots x_{N}$
becomes the eigenvalue density $\rho\left(x\right)$ for the ensemble.
Using the Sokhotski-Plemelj formula one can extract the eigenvalue
density $\rho\left(x\right)$ from the green function $G\left(z\right)$
as follows
\[
\rho\left(x\right)=\frac{1}{\pi}\lim_{\epsilon\rightarrow0^{+}}Im\left[G\left(x-i\epsilon\right)\right]
\]

In this work, we want to calculate properties of sums of random matrices
(the sum of the MCRM interactions and direct interactions). In general,
random matrices do not commute, and the spectrum of the sum matrix
isn't simply the sum of the spectra. Therefore it is hard to calculate
the spectrum of random matrices sum even given access to the Green
functions of the ensembles. Free probability is a tool generalizing
the notion of random variable independence to the field of random
matrices. Analogous to statistical independence for random variables,
two random matrix ensembles may exhibit the `freeness' property, the
precise definition can be found in \citep{livan_introduction_2018}.

Free probability provides us with a prescription for deriving the
Green function of the ensemble sum given the Green function of the
summed ensembles exhibiting the freeness property. This is analogous
to the convolution law for random variable sum. It proceeds as follows.
First, define the complex valued blue function to be the functional
inverse of the green function
\[
G\left(B\left(z\right)\right)=z
\]
Now, given the blue function of the two ensembles $B_{1}\left(z\right),B_{2}\left(z\right)$,
the blue function of the sum ensemble reads
\[
B\left(z\right)=B_{1}\left(z\right)+B_{2}\left(z\right)-\frac{1}{z}\ .
\]
Finally, to find the green function of the sum ensemble, invert the
blue function above using the relation
\[
B\left(G\left(z\right)\right)=z\ .
\]

\subsection{Wishart, GOE and Ginibre ensembles}

The perturbed MCRM interaction matrix appearing in Sec. \ref{subsec:Theory-for-the}
consists of a sum of two matrices:
\begin{enumerate}
\item Resource competition interaction matrix - Wishart ensemble
\[
\alpha_{ij}^{\left(r\right)}=\sum_{\beta=1}^{M}c_{i\beta}c_{j\beta}\negmedspace;\quad c_{i\beta}\sim Norm\left(\frac{\mu_{c}}{S},\frac{\sigma_{c}}{\sqrt{S}}\right)
\]
with the blue function 
\[
B_{W}\left(z\right)=\sigma_{c}^{2}\frac{\kappa}{1-\sigma_{c}^{2}z}+\frac{1}{z}\negmedspace;\quad\kappa=\frac{M}{N}\ .
\]
\item Direct competition interaction matrix - Ginibre ensemble
\[
\alpha_{ij}^{\left(d\right)}\sim Norm\left(\frac{\mu_{d}}{S},\frac{\sigma_{d}}{\sqrt{S}}\right)\ ;\enskip corr\left(\alpha_{ij}^{\left(d\right)},\alpha_{ji}^{\left(d\right)}\right)=\gamma\ .
\]
\end{enumerate}
In general (for $\gamma\ne1$) a matrix drawn from the Ginibre ensemble
is not Hermitian and therefore has a complex valued spectrum. Non
Hermitian ensembles call for a generalization of the Green function.
Concretely, in these cases the Green function would be a Quaternionic
valued function leading to much more complicated calculations. Luckily,
the Ginibre ensemble can be assembled as the sum of two independently
distributed matrices from the Gaussian orthogonal ensemble (GOE) with
complex prefactors \citep{burda_quaternionic_2015}. This representation
of the Ginibre ensemble allows for great simplification following
method by \citep{jarosz_novel_2004}.\\
Given two matrices $H,H'$, with elements drawn independently from
$H_{ij}\sim Norm\left(0,\sigma^{2}/N\right)$ and symmetrize $\left(H+H^{T}\right)/2$.
The Ginibre matrix $\alpha^{\left(d\right)}$ can be written as
\begin{align*}
\alpha^{\left(d\right)} & =c_{2}H+ic_{3}H'\\
c_{2} & =\sigma_{d}\omega\sqrt{1+\gamma}\\
c_{3} & =\sigma_{d}\omega\sqrt{1-\gamma}
\end{align*}
where $\omega$ is the aggression factor maintaining a constant direct
perturbation strength as described in \ref{subsec:Sensitivity-to-direct}.
The blue function of the GOE with real value prefactor $c$ is given
by
\[
B_{c\cdot GOE}\left(z\right)=\frac{1}{2}c^{2}z+\frac{1}{z}
\]

Finally, the ensemble for the perturbed interaction matrix $\alpha=\alpha^{\left(r\right)}+\omega\cdot\alpha^{\left(d\right)}$
can be written as

\begin{align*}
\alpha & =c_{1}W+c_{2}H+ic_{3}H'\\
c_{1} & =\sigma_{c}^{2}\\
c_{2} & =\sigma_{d}\omega\sqrt{1+\gamma}\\
c_{3} & =\sigma_{d}\omega\sqrt{1-\gamma}
\end{align*}
with the blue functions for the real and imaginary parts

\[
B_{\text{\ensuremath{\Re}}\alpha}\left(z\right)=B_{c_{1}W}+B_{c_{2}\cdot GOE}-\frac{1}{z}=c_{1}\frac{\kappa}{1-c_{1}z}+\frac{1}{2}c_{2}^{2}z+\frac{1}{z}
\]
\[
B_{\Im\alpha}\left(z\right)=\frac{1}{2}c_{3}^{2}z+\frac{1}{z}
\]

\subsection{Calculating the minimal eigenvalue of the matrix sum \label{subsec:Calculating-minimal-eigval}}

In this section we derive the minimal eigenvalue real part for the
ensemble describing the reduced interaction matrix of the perturbed
MCRM in Sec. \ref{subsec:Sensitivity-to-direct}. This is the main
result of this appendix, then being utilized to find the phase diagram
in Fig. \ref{fig:Phase diagram}.

In this section we follow the method by \citep{jarosz_novel_2004}
to find the spectrum of the sum of Hermitian random matrices with
imagery prefactors. Using this method one can derive the spectrum
of a non-Hermitian random matrix $H_{1}+iH_{2}$ comprised of two
Hermitian matrices $H_{1},H_{2}$, without having to go through cumbersome
calculations green functions quaternionic.

Still, it is hard to find the entire spectrum of this ensemble for
$\alpha$. We simplify the problem further, trying to find just the
minimal real part of the complex spectrum, given as a particular case
of the equations for spectrum support contour on the complex plane.

\[
g\equiv a+ib\ ;\qquad g^{I}=\alpha+i\beta
\]
By \citep{jarosz_novel_2004}, Eq. (77),

\[
B_{H}\left(g\right)=\frac{c_{1}\kappa}{1-c_{1}g}+\frac{1}{2}c_{2}^{2}g+\frac{1}{g}
\]
\[
gB_{H}\left(g\right)=\frac{1}{2}c_{2}^{2}\left(a^{2}-b^{2}+2iab\right)+c_{1}\kappa\frac{\left(1-c_{1}a\right)a-c_{1}b^{2}+ib}{\left(1-c_{1}a\right)^{2}+c_{1}^{2}b^{2}}+1
\]
\[
\frac{gB_{H}\left(g\right)-\bar{g}B_{H}\left(\bar{g}\right)}{g-\bar{g}}=c_{2}^{2}a+c_{1}\kappa\frac{1}{\left(1-c_{1}a\right)^{2}+c_{1}^{2}b^{2}}
\]
\[
\Downarrow
\]
\[
x=c_{2}^{2}a+\frac{c_{1}\kappa}{\left(1-c_{1}a\right)^{2}+c_{1}^{2}b^{2}}
\]
By \citep{jarosz_novel_2004}, Eq. (78),

\[
B_{H'}\left(g^{I}\right)=\frac{1}{2}c_{3}^{2}g^{I}+\frac{1}{g^{I}}
\]
\[
g^{I}B_{H'}\left(g^{I}\right)=\frac{1}{2}c_{3}^{2}\left(g^{I}\right)^{2}+1=\frac{1}{2}c_{3}^{2}\left(\alpha^{2}-\beta^{2}+2i\alpha\beta\right)+1
\]
\[
g^{I}B_{H'}\left(g^{I}\right)-\bar{g}^{I}B_{H'}\left(\bar{g}^{I}\right)=2ic_{3}^{2}\alpha\beta
\]
\[
g^{I}B_{H'}\left(g^{I}\right)-\bar{g}^{I}B_{H'}\left(\bar{g}^{I}\right)=y\left(g^{I}-\bar{g}^{I}\right)
\]
\[
\Downarrow
\]
\[
y=\frac{\alpha}{c_{3}^{2}}
\]
And from \citep{jarosz_novel_2004}, Eq. (74),
\[
\frac{B_{H}\left(g\right)-B_{H}\left(\bar{g}\right)}{g-\bar{g}}+\frac{B_{H'}\left(g^{I}\right)-B_{H'}\left(\bar{g}^{I}\right)}{g^{I}-\bar{g}^{I}}+\frac{1}{g\bar{g}}=0
\]
\begin{align*}
B_{H}\left(g\right) & =\frac{c_{1}\kappa}{1-c_{1}g}+\frac{1}{2}c_{2}^{2}g+\frac{1}{g}\\
 & =c_{1}\kappa\frac{1-c_{1}a+c_{1}ib}{\left(1-c_{1}a\right)^{2}+c_{1}^{2}b^{2}}+\frac{1}{2}c_{2}^{2}g+\frac{\bar{g}}{g\bar{g}}
\end{align*}
\[
\frac{B_{H}\left(g\right)-B_{H}\left(\bar{g}\right)}{g-\bar{g}}=c_{1}\kappa\frac{c_{1}}{\left(1-c_{1}a\right)^{2}+c_{1}^{2}b^{2}}+\frac{1}{2}c_{2}^{2}-\frac{1}{a^{2}+b^{2}}
\]
Using \citep{jarosz_novel_2004}, Eq. (63) we have $g\bar{g}=g^{I}\bar{g}^{I}$

\[
\frac{B_{H'}\left(g^{I}\right)-B_{H'}\left(\bar{g}^{I}\right)}{g^{I}-\bar{g}^{I}}=\frac{1}{2}c_{3}^{2}-\frac{1}{\alpha^{2}+\beta^{2}}=\frac{1}{2}c_{3}^{2}-\frac{1}{a^{2}+b^{2}}
\]
\[
\Downarrow
\]
\[
\frac{c_{1}^{2}\kappa}{\left(1-c_{1}a\right)^{2}+c_{1}^{2}b^{2}}+\frac{1}{2}\left(c_{2}^{2}+c_{3}^{2}\right)-\frac{1}{a^{2}+b^{2}}=0
\]
Combining \citep{jarosz_novel_2004}, Eqs. (63,74,77) we get the set
of coupled equations

\[
\begin{cases}
x=c_{2}^{2}a+\frac{c_{1}\kappa}{\left(1-c_{1}a\right)^{2}+c_{1}^{2}b^{2}}\\
\frac{c_{1}^{2}\kappa}{\left(1-c_{1}a\right)^{2}+c_{1}^{2}b^{2}}+\frac{1}{2}\left(c_{2}^{2}+c_{3}^{2}\right)-\frac{1}{a^{2}+b^{2}}=0
\end{cases}
\]
According to \citep{jarosz_novel_2004}, Eq. (93) the spectrum contour
equation is given by $\left(g+\bar{g}\right)^{2}+\left(g^{I}+\bar{g}^{I}\right)^{2}=4g\bar{g}$.
Now, focusing on the real part of the contour, given by $y=0$ combined
with \citep{jarosz_novel_2004}, Eq. (78) leads to $\alpha=0$. Therefore
the contour equation reduce to 
\[
\left(g+\bar{g}\right)^{2}=4g\bar{g}\Rightarrow4a^{2}=4\left(a^{2}+b^{2}\right)\Rightarrow b^{2}=0
\]
Plugging that back to the set of coupled equations above, yields the
polynomial equation for $a$ 

\begin{align}
x & =c_{2}^{2}a+\frac{c_{1}\kappa}{\left(1-c_{1}a\right)^{2}}\label{eq:eigenvalue_contour_real}\\
0 & =c_{1}^{2}\left(c_{2}^{2}+c_{3}^{2}\right)a^{4}-2c_{1}\left(c_{2}^{2}+c_{3}^{2}\right)a^{3}+\nonumber \\
 & +\left(2\left(\kappa-1\right)c_{1}^{2}+c_{2}^{2}+c_{3}^{2}\right)a^{2}+4c_{1}a-2\nonumber 
\end{align}

Substituting back the real $a$ roots into Eq. (\ref{eq:eigenvalue_contour_real})
to get the minimal and maximal eigenvalue real parts of the ensemble.
By doing so determining the linear stability of the perturbed MCRM
model at Eq. (\ref{eq:Reduced MCRM EOM}).

\section{Species abundance distribution\label{sec:Species-abundance-distribution}}

At fixed points of resource-consumer models, species diversity is
limited by the number of limiting resources ($S^{*}\leq M$), according
to the competitive exclusion principle \citep{mcgehee_mathematical_1977}.
In contrast, non-equilibrium states are not bound by the exclusion
principle and can exceed this limit, i.e. $M<S^{*}$. In this section
we show that for the system described in Eq. (\ref{eq:Tanh MCRM EOM})
this is indeed the case, even if the migration is very small (in the
$\eta\rightarrow0$ limit).

As discussed in Sec. \ref{subsec:Intra-community-=000026-diversity},
simulations with migration $\eta$, show a typical species abundance
probability distribution, see Fig. \ref{fig:diversity-above-comp-ex}(A),
with a power law as the abundance distribution between an upper value
$N_{u}$, and lower value determined by the migration floor $\eta$.
We write this probability distribution as

\[
P\left(N\right)=\begin{cases}
P_{high}\left(N\right) & N_{u}<N\\
cN^{-\left(\nu+1\right)} & b\eta<N<N_{u}\\
P_{low}\left(N\right) & \eta\lesssim N<b\eta
\end{cases}
\]
The power law behavior is parameterized by $c,\nu$ that may dependent
on the migration $\eta$. The lower part $P_{low}$$\left(N\right)$
is the abundance distribution for species that are maintained thanks
to migration. We define this region to go up to $b\eta$ with a (somewhat
arbitrary) constant value $b$.

Our interest is in when and how species diversity goes beyond the
competitive exclusion bound in this probability distribution. Let
us define $C_{high},N_{CE},C_{CE}$ as

\[
\intop_{N_{u}}^{\infty}P\left(N\right)dN\equiv C_{high}\ ,
\]
\[
\intop_{N_{CE}}^{\infty}P\left(N\right)dN=\frac{M}{S}\ ,
\]
\[
\intop_{b\eta}^{N_{CE}}P\left(N\right)dN\equiv C_{CE}\ .
\]
Note that if competitive exclusion holds and $\eta=0$, there are
no species in the range $0<N<N_{CE}$. That is, species are either
extinct and concentrated at $N=0$ of $P\left(N\right)$, or have
an abundance above $N_{CE}$. Therefore, to demonstrate that the species
diversity can exceed the competitive exclusion limit in chaotic states
we show that $0<C_{CE}$ at the limit of vanishing migration.

For simplicity of the analysis we replace $P_{low}$ by extending
the power law and introducing a new lower cutoff at $a\eta$, to preserve
the area under the curve of this lower region. We still treat abundances
below $b\eta$ as species maintained solely by migration. The simplified
probability distribution reads

\[
P\left(N\right)=\begin{cases}
P_{high}\left(N\right) & N_{u}<N\\
cN^{-\left(1+\nu\right)} & a\eta<N<N_{u}
\end{cases}\ .
\]
From normalization of $P\left(N\right)$ we solve for $\frac{c}{\nu}$
to find

\[
1-C_{up}=\intop_{a\eta}^{N_{u}}cN^{-\left(1+\nu\right)}dN=-\frac{c}{\nu}\left[N_{u}^{-\nu}-\left(a\eta\right)^{-\nu}\right]\ ,
\]
\[
\frac{c}{\nu}=\frac{\left(1-C_{up}\right)}{\left(a\eta\right)^{-\nu}-N_{u}^{-\nu}}\ .
\]
Next, we express $N_{CE}$ as

\[
\frac{M}{S}-C_{up}=\intop_{N_{CE}}^{N_{u}}cN^{-\left(1+\nu\right)}dN=-\frac{c}{\nu}\left[N_{u}^{-\nu}-N_{CE}^{-\nu}\right]
\]
\[
N_{CE}^{-\nu}=\left(a\eta\right)^{-\nu}\left(\frac{\frac{M}{S}-C_{up}}{1-C_{up}}\right)+N_{u}^{-\nu}\left(\frac{1-\frac{M}{S}}{1-C_{up}}\right)
\]
Finally, $C_{CE}$ takes the form

\[
C_{CE}=\intop_{b\eta}^{N_{CE}}\tilde{c}N^{-\left(1+\nu\right)}dN=-\frac{c}{\nu}\left[N_{CE}^{-\nu}-\left(b\eta\right)^{-\nu}\right]
\]
\[
C_{CE}=\frac{\left(b\eta\right)^{-\nu}-\left(a\eta\right)^{-\nu}}{\left(a\eta\right)^{-\nu}-N_{u}^{-\nu}}\left(1-C_{up}\right)+\left(1-\frac{M}{S}\right)
\]
At the limit $\eta\rightarrow0$
\[
C_{CE}=\left(\frac{a}{b}\right)^{\nu}\left(1-C_{up}\right)+C_{up}-\frac{M}{S}
\]
This equation expresses $C_{CE}$ as a function of the parameters
of the probability distribution ($a,b,\nu,C_{up}$), and the number
of species $M$ and resources $S$. Simulations shows that $0\leq\nu\ll1$
(possibly vanishing) at the limit $\eta\rightarrow0$. Assuming that
$M/S<1$ (there are more species in the pool than resources), we conclude
that $0<C_{CE}$, hence non-equilibrium states of consumer-resource
models can sustain diversity exceeding the competitive exclusion limit.

\begin{figure}
\begin{centering}
\includegraphics[width=1\columnwidth]{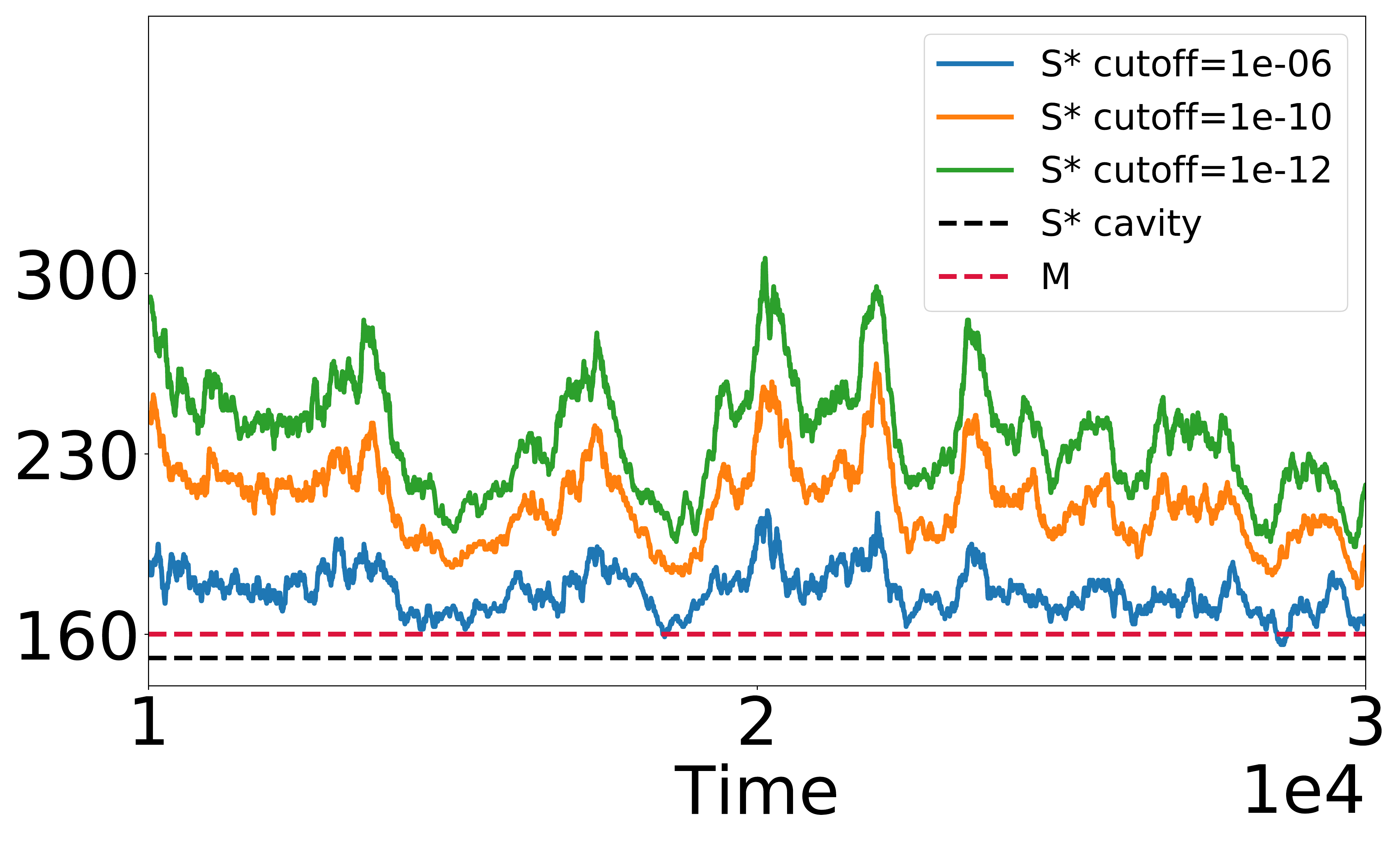}
\par\end{centering}
\caption{\label{fig:s_star_vs_cavity}Number of coexisting species $S^{*}$,
as a function of time in the persistent dynamics phase. A species
is counted in the standing diversity $S^{*}$ if its abundance is
above the level given in the legend. As can be seen, the number coexisting
species exceeded the amount of resources for large range of abundance
levels. }
\end{figure}

Finally, a note regarding the diversity, compared with the diversity
as predicted by the cavity solution described in Appendix \ref{sec:Cavity-equations}.
That solution is only exact when the system reaches a unique stable
equilibrium. Elsewhere it is an approximation; from simulations of
the first model variant in Sec. \ref{subsec:Sensitivity-to-direct},
we find that the cavity solution is lower than the one described in
this Section, see Fig. \ref{fig:s_star_vs_cavity}.

\section{Stiff and soft fluctuation directions\label{sec:Stiff-and-soft}}

It is interesting to see whether the fluctuations of the abundances
are directly related to the marginal directions of the MCRM fixed
points. To do that, we rotate the vector $\left\{ N_{i}\right\} _{i=1}^{S}$
in a way that will separate the ``stiff'' degrees of freedom, lying
in the non-marginal directions of a MCRM fixed point, and the ``soft''
degrees of freedom at the marginal dimensions. This is done by rotating
with an orthogonal matrix $O$ the abundance vector $\vec{N}\left(t\right)$
from a simulation of the perturbed MCRM in it's chaotic phase. The
orthogonal matrix $O$ is obtained from the spectral decomposition
of the unperturbed interaction matrix $\alpha_{ij}^{\left(r\right)}=\sum_{k,l}O_{ik}D_{kl}O_{jl}$
where $D$ is diagonal matrix. Note that since $\alpha^{\left(r\right)}$
is a symmetric positive-semi-definite matrix, its eigenvalues $\left\{ \lambda_{i}\right\} _{i=1}^{S}$
are real valued and non-negative.

\begin{figure}
\begin{centering}
\includegraphics[width=0.95\columnwidth]{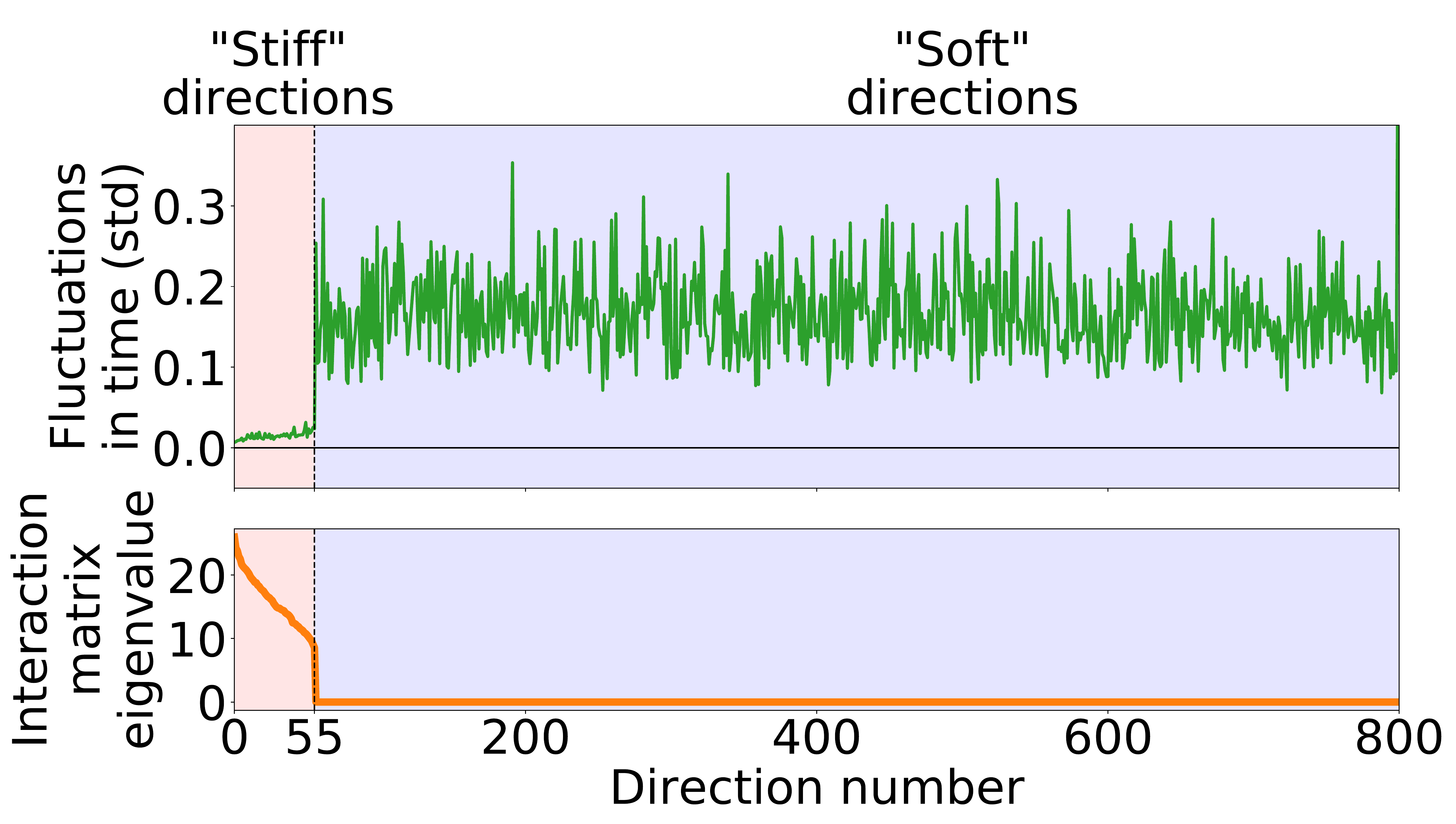}
\par\end{centering}
\caption{(Top) Fluctuations over time along the eigenvector directions of the
resource interaction matrix $\alpha^{\left(r\right)}$. (Bottom) Corresponding
eigenvalues $\lambda_{i}$ of the spectral decomposition of $\alpha^{\left(r\right)}$.
There is a clear distinction between ``Stiff'' directions showing
little fluctuations corresponding to finite positive eigenvalues,
and ``Soft'', strongly fluctuating marginal directions with corresponding
zero eigenvalues.\label{fig:fluctuations_vs_directions}}
\end{figure}

Denote by $y_{i}\left(t\right)=\sum_{j=1}^{S}O_{ji}N_{i}\left(t\right)$
the rotated degrees of freedom. These are a combination of species
abundances at time $t$. Define the fluctuation over time in direction
$i$ to be $\std\left(y_{i}\right)\equiv\sqrt{\left\langle y_{i}\left(t\right)-\left\langle y_{i}\left(t\right)\right\rangle _{t}\right\rangle _{t}}$,
where $\left\langle ...\right\rangle _{t}$ denotes time average over
a $\Delta t=10^{4}$ interval . Plotting $\std\left(y_{i}\right)$
sorted by the eigenvalue $\lambda_{i}$, shows that the fluctuations
in the ``soft'' directions (where $\dot{\lambda_{i}=0}$) are consistently
and significantly larger than in the ``stiff'' directions (where
$\lambda_{i}>0$), see Fig. \ref{fig:fluctuations_vs_directions}.

\section{Isolated systems (no migration)\label{sec:Meta-community}}

Here we discuss cases where there is no migration from an external
``mainland'' pool of species. We consider both a single community,
and a a meta-community, a setting in which multiple well-mixed communities
are coupled by migration. We show that meta-communities can allow
for persistent dynamics over long times, even in the absence of external
migration, and for finite population sizes. In isolated well-mixed
communities, simulations show that extinctions drive the system to
a fixed point, with diversity a little below the competitive exclusion
bound.

The dynamics of the meta-community are a set of differential equations
for $N_{i}^{\left(u\right)}$ describing the abundance of the $i$-th
species in the $u$-th community,
\[
\frac{dN_{i}^{\left(u\right)}}{dt}=...+\sum D_{i}^{\left(u,v\right)}\left[N_{i}^{\left(v\right)}-N_{i}^{\left(u\right)}\right]\ ,
\]
where the ``...'' refers to the terms in the RHS of Eq. (\ref{eq:Reduced MCRM EOM})
applied to $N_{i}^{\left(u\right)}$, with $\eta_{i}=0$. A species
is considered extinct and removed from the system when its abundance
$N_{i}^{\left(u\right)}$ goes below some cut-off $N_{c}$ in all
communities $u$, corresponding to the inverse of the population size.

\begin{figure}
\begin{centering}
\includegraphics[width=1\columnwidth]{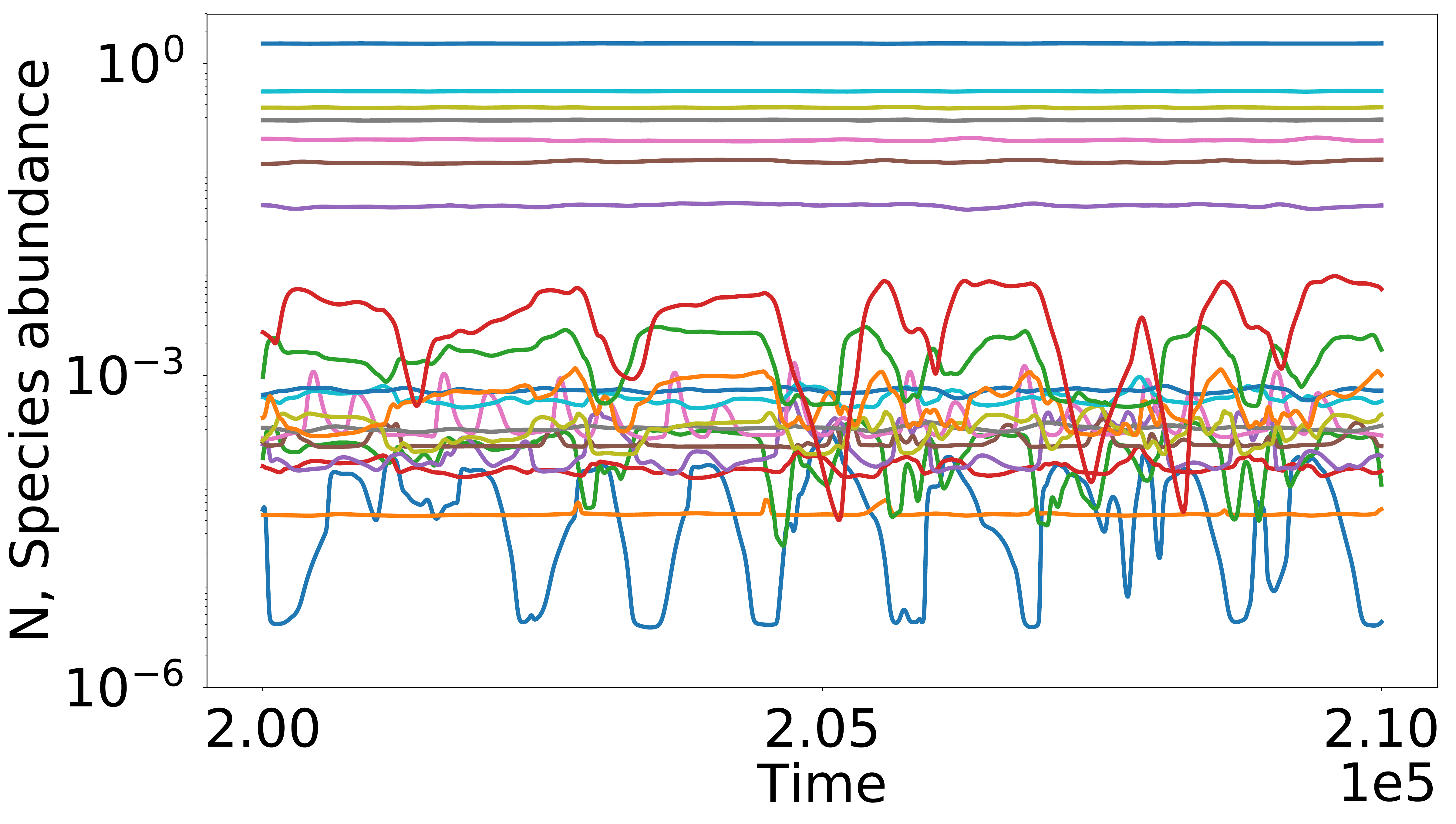}
\par\end{centering}
\caption{Dynamics of a meta-community composed of $8$ communities coupled
by migration, at late times. Persistent abundance fluctuations are
shown, which do not go below some value, showing that even finite
populations can persist for very long times. 20 representative species
are plotted. \label{fig:meta-community}}
\end{figure}

Fig. \ref{fig:meta-community} shows the dynamics at late times of
a few representative abundances $N_{i}^{\left(u\right)}$, showing
persistent fluctuations in a meta-community comprised of $8$ communities
with $S=400$ species and $M=80$ resources. The model in each patch
corresponds to that in Sec. \ref{subsec:Sensitivity-to-direct}. The
resource interaction matrix $\alpha^{\left(r\right)}$ has $\mu_{c}=30$
and $\sigma_{c}=6$. The matrix $\alpha$ is very similar but not
identical between the different communities, with correlation $\rho=0.997$
between the $\alpha_{ij}=\alpha_{ij}^{\left(r\right)}+\omega\alpha_{ij}^{\left(d\right)}$
for the same $i,j$ in different communities. Direct interaction matrix
$\alpha^{\left(d\right)}$ is drawn independently for each community
with $\mu_{d}=10$, $\sigma_{d}=20$ and $\gamma=0$. As in the main
text, $\omega$ is determined to satisfy perturbation strength of
$0.05$. Coupling between communities set to be $D=10^{-4}$. Cutoff
abundance is taken to be $N_{c}=10^{-20}$.

\begin{figure}
\begin{centering}
\includegraphics[width=1\columnwidth]{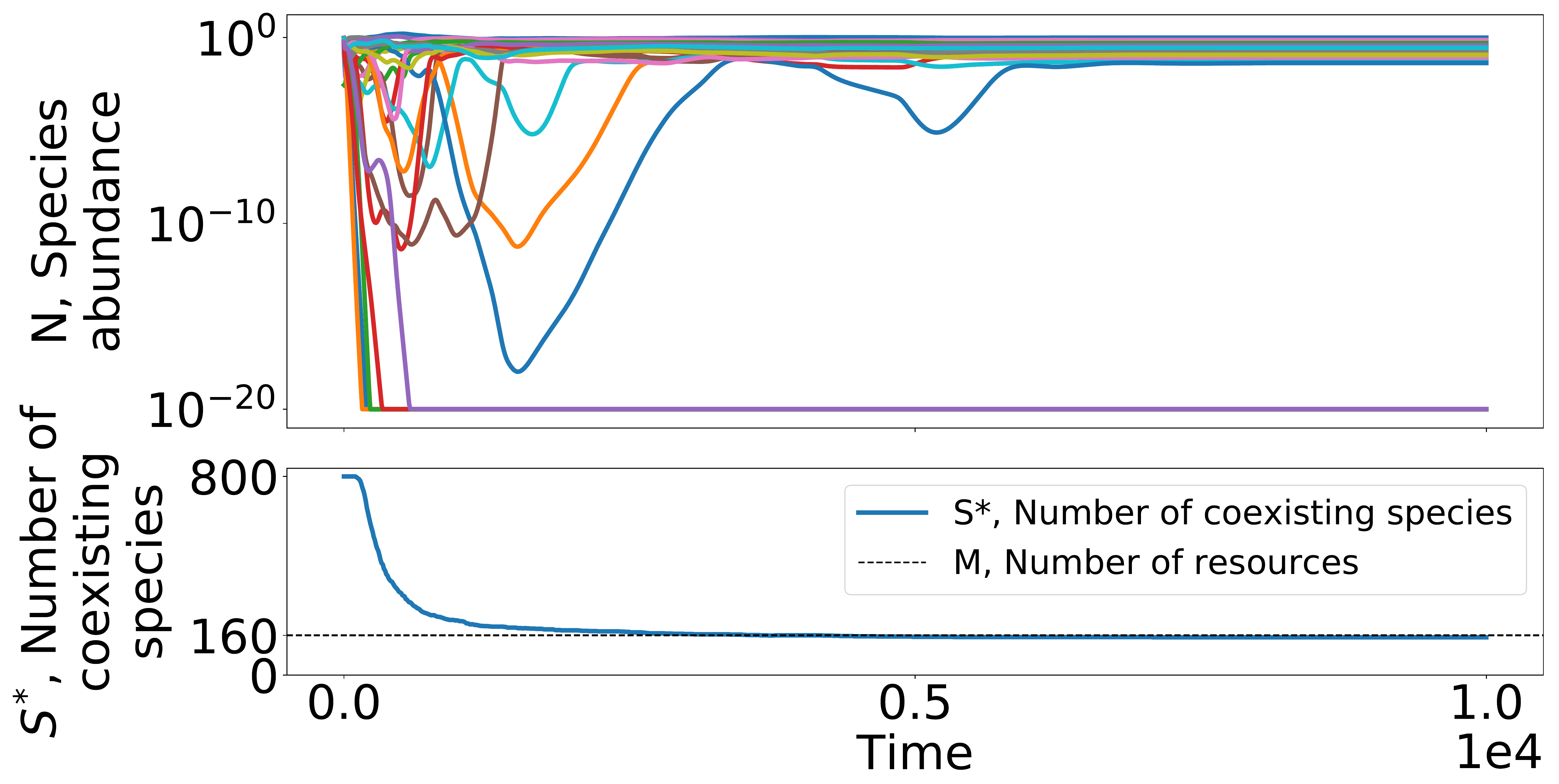}
\par\end{centering}
\caption{\label{fig:dynamics-no-migration}Dynamics of a single community without
external migration at the chaotic phase. Species abundances initially
fluctuate, and some go extinct. A fixed point is reached once the
diversity goes a little below the number of resources $M$.}
\end{figure}

A simulation of a single, well-mixed community is shown in Fig. \ref{fig:dynamics-no-migration},
along with the diversity as a function of time. The diversity drops,
until the system reaches a fixed point with $S^{*}$ a little below
$M$, almost saturating the competitive exclusion bound. Simulation
parameters are specified in Tab. \ref{tab:simulation_parameters}
and are similar to that in Fig. \ref{fig:Summary-of-argument}(C)
but with $\eta=0$.

\section{Symmetric additional interactions\label{sec:Symmetric-additional-interaction}}

Here we consider a setting similar to that in Sec. \ref{subsec:Sensitivity-to-direct},
where additional interactions $\alpha_{ij}^{\left(d\right)}$ are
added to the ground model. The difference is that here they are taken
to be symmetric, $\alpha_{ij}^{\left(d\right)}=\alpha_{ji}^{\left(d\right)}$.
This difference is important, since in this case, the entire interaction
matrix $\alpha$ is symmetric. This means that the dynamics admit
a Lyapunov function, and always reach a fixed point. A similar situation,
with symmetric random Lotka-Volterra interactions (in this work's
terminology, $\alpha=\alpha^{\left(d\right)}$) has been studied in
\citep{biroli_marginally_2018}. There, a fixed point phase was found.
Beyond it lies a \emph{critical} phase, characterized by many alternative
equilibria, all of them close to marginal stability, namely such that
the minimal eigenvalue $\lambda_{min}\rightarrow0$ as $S\rightarrow\infty$.
Specifically, it was found that $\lambda_{min}\propto S^{-2/3}$.

\begin{figure}
\begin{centering}
\includegraphics[width=1\columnwidth]{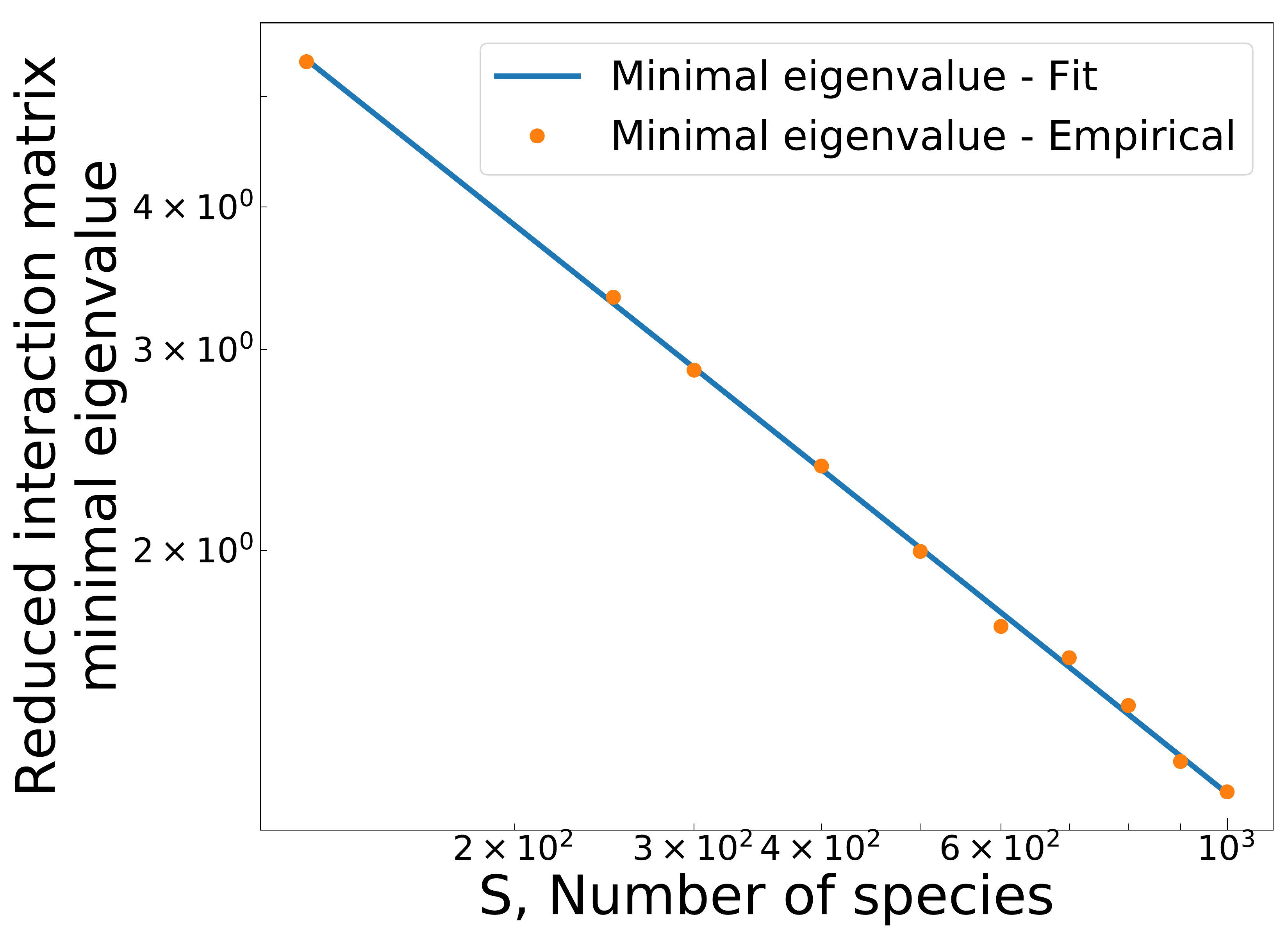}
\par\end{centering}
\caption{\label{fig:min_eigval_gamma_one-fit}Fitting the minimal eigenvalue
of the reduced interaction matrix to $\lambda_{min}\left(S\right)=a\cdot S^{b}+\lambda_{\infty}$.
Parameters as in Fig. \ref{fig:min_eigval_gamma_one}, with $\sigma_{c}=23$.}
\end{figure}

\begin{figure}
\begin{centering}
\includegraphics[width=1\columnwidth]{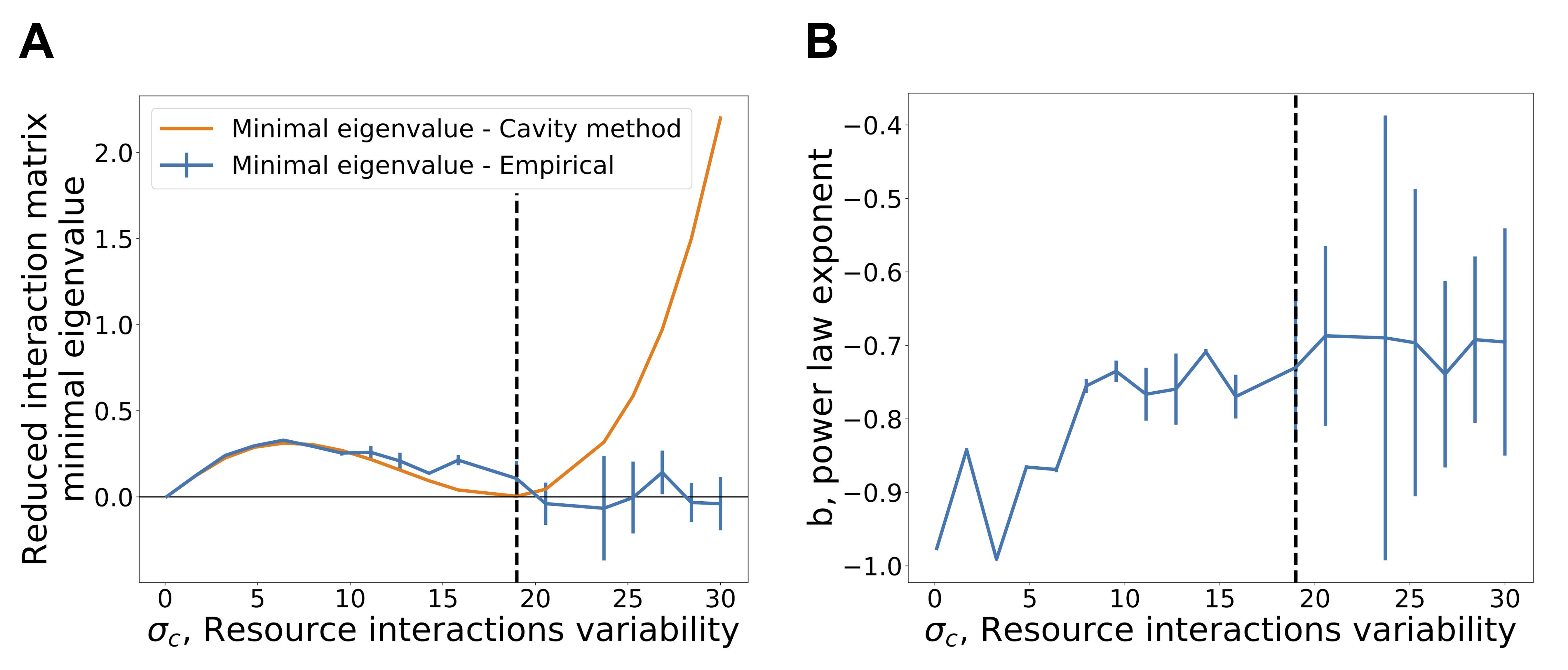}
\par\end{centering}
\caption{\label{fig:min_eigval_gamma_one}Minimal eigenvalue of the reduced
interaction matrix with symmetric direct interactions perturbation
($\gamma=1$). Minimal eigenvalue at $S\rightarrow\infty$ is obtained
from a fit to $\lambda_{min}\left(S\right)=a\cdot S^{b}+\lambda_{\infty}$,
see Fig. \ref{fig:min_eigval_gamma_one-fit}. Shown are (A) The minimal
eigenvalue $\lambda_{\infty}$, and (B) the power law exponent $b$.}
\end{figure}

Here we find precisely the same phenomenology, with a fixed point
phase. Beyond it simulations show that the system possesses multiple
alternative equilibria. Furthermore, the minimal eigenvalue was measured
for multiple values of $\sigma_{c}$ and $S$, and averaged over many
runs. For each value of $\sigma_{c}$ it was fit to $\lambda_{min}\left(S\right)=a\cdot S^{b}+\lambda_{\infty}$,
where $a,b,\lambda_{\infty}$ depend on $\sigma_{c}$, see Fig. \ref{fig:min_eigval_gamma_one-fit}.
The results for $b,\lambda_{\infty}$ are shown in Fig. \ref{fig:min_eigval_gamma_one}.
We find that beyond the unique fixed point phase the results are very
different from the simple cavity solution for this case, and consistent
with $\lambda_{\infty}=0$ and $b=-2/3$ which was predicted for the
random Lotka-Volterra setting.

\end{document}